\documentclass[%
 reprint,
onecolumn,
 amsmath,amssymb,
 aps,
]{revtex4-1}

\usepackage{graphicx}% Include figure files
\usepackage{dcolumn}% Align table columns on decimal point
\usepackage{bm}% bold math
\usepackage{subfigure}
\usepackage{color}
\usepackage{mathrsfs}

\usepackage{todonotes}
\usepackage{float} 
\usepackage{braket}
\usepackage{tikz}
\usetikzlibrary{positioning, shapes.geometric}
\usetikzlibrary{calc,positioning,decorations.pathreplacing,calligraphy,matrix,graphs,fit,shapes,shapes.geometric,arrows.meta,patterns}
\begin{document}

\preprint{APS/123-QED}

\title{\boldmath Quantum correlations of a two-qubit system and the Aubry-Andr\'{e} chain in bosonic environments}% Force line breaks with \\

\author{He Wang}
\affiliation{College of Physics, Jilin University,\\Changchun 130021, China}%
\affiliation{State Key Laboratory of Electroanalytical Chemistry, Changchun Institute of Applied Chemistry,\\Changchun 130021, China}%
\author{Liufang Xu}
\email{lfxuphy@jlu.edu.cn}
\affiliation{College of Physics, Jilin University,\\Changchun 130021, China}%
\author{Jin Wang}
\email{jin.wang.1@stonybrook.edu}
\affiliation{Department of Chemistry and of Physics and Astronomy, Stony Brook University, Stony Brook,\\NY 11794-3400, USA}%

\date{\today}% It is always \today, today,
\begin{abstract}

In this research, we analyze two models using the tensor network algorithm. The quantum correlations of a two-qubit system are first studied in different bosonic reservoirs. Both equilibrium and nonequilibrium scenarios are discussed. Non-Markovian effects can improve the survival time of the quantum correlations significantly and weaken the decoherence effect. Non-Markovian dynamics with existing memory can lead to entanglement rebirth in specific scenarios instead of the eventual entanglement decay or death seen in memoryless Markovian cases. The system reaches a steady state quickest in sub-Ohmic reservoirs and shows the most apparent non-Markovian behavior in super-Ohmic reservoirs. %At fixed times, we find that memory can boost the correlations. However, more memory is not always better, and too much memory can also cause decoherence. 
The Markovian approximation used in this paper is superior to that in the Bloch--Redfield master equation. The entanglement dynamics behave similarly under different approaches when the system-bath coupling is weak, and the memory effect is significant when the system-bath coupling is strong. %We also examine how to confront decoherence from the environment in practical teleportation. The decay rates of the entanglement and the fidelity of quantum teleportation can be slowed down significantly by regulating the external field.
We not only study the impact of the environment on quantum correlations, but also how to protect quantum correlations. Starting from a state in which the two ends are maximally entangled, a one-dimensional Aubry-Andr\'{e} chain model is also studied. We identify distinct phases by monitoring the imbalance dynamics. When the chain is closed, the imbalance dynamics behave differently in various phases, and so does the entanglement evolution between the chain's ends. When the first site couples to a bath, we found the imbalance dynamics can still be an effective indicator to differentiate various phases in an early evolution stage since the imbalance dynamics is only remarkably affected at relatively high temperatures. The distribution of the eigenenergy of the system can account for it. The entanglement of the chain ends decays rapidly in all phases due to one of the ends being coupled to the bath directly. However, the entanglement of the chain ends will persist for a perceptible amount of time in the localization phase if the bath is coupled to the middle site of the chain. Our research shows that one can utilize the disordered environment as a buffer to protect quantum correlations.
\end{abstract}

\maketitle

\section{Introduction\label{Introduction}}

In recent years, theoretical and experimental progress has demonstrated that quantum correlations (including coherence \cite{Quantifying_Coherence}, entanglement \cite{Quantum_entanglement}, and quantum discord \cite{Quantum_Discord}) are essential resources for quantum information processing, including quantum teleportation \cite{Teleporting}, quantum cryptography \cite{Quantum_Cryptography}, and quantum dense coding \cite{Quantum_Dense_Coding}. Quantum correlations are not only the primary resources for quantum information but also the cornerstone of the realization of quantum communication and quantum computing \cite{Quantum_Computation_and_Quantum_Information}. A physical system inevitably couples to the environment, which can cause decoherence that can eliminate the quantum correlations. However, environments can restore quantum correlations or preserve them under certain conditions \cite{Controlling_entanglement_generation_in_external_quantum_fields,Thermal_amplification_of_field-correlation_harvesting,Two_Accelerated_Detectors}. Therefore, the study of quantum correlations in an open system can help determine appropriate environments for the implementation of quantum computing or other quantum information processing.

Open quantum system theory plays a crucial role in modern quantum mechanics. When the coupling between the system and environment is weak, the memory time of the environment is short compared to the time scale of the evolution of the system, and the Born--Markov approximation is applicable \cite{The_Theory_of_Open_Quantum_Systems}. Although this approximation is very effective within its area of applicability, for most real systems in the strong system-bath coupling regime or in environments with long-duration correlations, the approximation is rather limited and not justified. For instance, the behaviors of superconducting qubits in circuit QED systems \cite{C-QED}, NV centers in diamonds \cite{NV}, and quantum dots in semiconductors \cite{Quantum_Dot} all require a strong coupling description. It has been experimentally demonstrated that non-Markovian behaviors inevitably emerge in such strong system-bath coupling systems \cite{Experimental_non-Markovian,Global_correlation_and_local_information_flows_in_controllable_non-Markovian_open_quantum_dynamics}. The corresponding theoretical approaches go beyond the Markovian regime to the non-Markovian regime, which includes the Nakajima–Zwanzig projection operator equations \cite{The_Theory_of_Open_Quantum_Systems}, time-convolutionless master equations \cite{The_Theory_of_Open_Quantum_Systems}, Keldysh--Lindblad equations \cite{Keldysh_diagrams_perturbation_theory}, and reaction co-ordinate methods \cite{reaction_co-ordinate}. However, these approaches are also limited to the weak system-bath coupling regimes. Approaches that hold for the strong system-bath coupling regime include the polaron transformed master equation \cite{polaron1,polaron2}, the hierarchal equations of motion (HEOM) method \cite{HEOM}, and the influence functional (IF) method \cite{IF1,IF2,IF3,IF4}. We mainly focus on the IF method in this paper.

The IF method integrates all the influences from the environment \cite{IF1}. However, the cost of computing the IF without any approximations is huge, and the size of the IF scales exponentially with the number of time steps. With a finite memory approximation, Makri and Makarov showed that the path integral can be reformulated as a propagator of the augmented density tensor (ADT) that encodes the system’s history \cite{IF2,IF3}. The IF can be assembled as a series of influence functions, where an influence function quantifies how the evolution of the system at some time is influenced by the state of the system at an earlier time. This approach is called the Quasi-Adiabatic Path Integral (QUAPI) \cite{IF2,IF3}. Naturally, the IF can be described by a matrix product operator (MPO), and the ADT can be efficiently represented and propagated in the form of a matrix product state (MPS) in the tensor network language \cite{tn1,tn2,TEMPO1,TEMPO2,TEMPO3}. The resulting time-evolving matrix product operators (TEMPO) method is numerically exact. It has been widely used in many studies, including for the optimal control of non-Markovian open quantum systems \cite{optimal_control}, non-additive effects of environments \cite{nonadditive}, quantum heat statistics \cite{heat_statistics}, and the thermalization of a one-dimensional many-body system \cite{thermalization}. The TEMPO can be recast in the process tensor (PT) frame \cite{pt1}, where the PT is a multi-linear map from the set of all possible control operation sequences in the lab on the system to the resulting output states, and it can generally be expressed in MPO form \cite{pt2}. The process of constructing the tensor network IF for general dynamics can be found in \cite{pt3}.

It is of interest to us to investigate the behaviors of the quantum correlations within systems strongly coupled with various environments and to understand how to regulate and control them. On the one hand, this is important for the theoretical study of open quantum system theory and quantum information science. On the other hand, it can provide technical support in practical scenarios. For example, the decoherence and disentanglement of an open system in non-Markovian environments are apparently distinct from those in Markovian environments due to the backflow of information. The memory effect of non-Markovian dynamics may preserve the quantum correlations over a more extended period, which opens up the potential for realizing quantum technologies \cite{Non-Markovian_entanglement_preservation,Continuous_variable_entanglement_dynamics_in_structured_reservoirs,Entanglement_oscillations_in_non-Markovian_quantum_channels}. It has been shown that engineering a structured non-Markovian environment is also meaningful in protecting the system from decoherence \cite{cheng2016preservation,mu2018microscopic}. 
In this paper, we utilize the process tensor-time evolving block decimation (PT-TEBD) algorithm \cite{thermalization,TEMPO3,TEBD} to study the quantum correlations of a two-qubit system strongly coupled to a bosonic environment in both equilibrium and nonequilibrium scenarios. The non-Markovianness influences the dynamics of the correlations remarkably in all types of baths. The entanglement may undergo rebirth after sudden death due to information backflow rather than the eventual death seen in the dynamics under memoryless conditions. Oscillations of the correlations are common in the different types of baths. The amplitudes of the correlation dynamics of the two qubits in super-Ohmic reservoirs are the largest and decay the slowest, showing the strongest memory effect, while those in sub-Ohmic reservoirs reach a steady value the quickest. The higher the temperature, the faster the system arrives at a steady state. %We find that a reservoir with memory can boost the correlations, but more memory is not always better, and too much memory can sometimes also destroy the quantum correlations. 
We also study the Markovian approximation used in this paper, under which we keep only the influence of the last one-step history of the system. This approximation works well when the system-bath coupling is weak and is superior to the Bloch--Redfield master equation. The memory leads to entanglement revival in the strong system-bath coupling regime. Thus, one can anticipate that only the whole memory can provide an exact prediction when the system-bath coupling is strong. 
%To overcome the decoherence effect from the baths, we apply environmental engineering to the quantum teleportation protocol. By introducing a controlled external field, the decay rates of the entanglement and the fidelity of quantum teleportation are lowered significantly.

The protection of entanglement is another crucial subject. It has been suggested that introducing disorder into the environment might help to prevent thermalization and preserve entanglement \cite{wang2018time}. Inspired by this, we investigate the quasi-disordered (AA) Aubry-Andr\'{e} chain coupled with the environment. The AA model has abundant phases of matter in various parameter zones, i.e., ergodic, MBL (many-body localization), and AL (Anderson localization) phases. The AL is where the idea for the MBL originated. The localization of non-interacting particles in a disordered system is referred to as the AL, and the particle localization is completely caused by disordered external potentials \cite{anderson1958absence}. Besides the completely disordered external potentials, the quasi-periodic external potentials can also cause AL. The AA chain is the well-known model in such a study. The AL phenomena have been shown by experiments in both completely disordered systems \cite{billy2008direct} and quasi-periodic systems \cite{roati2008anderson}. 
More recently, once interactions are incorporated, such systems were shown to exhibit MBL \cite{gornyi2005interacting, basko2006metal, altman2015universal, nandkishore2015many, alet2018many, abanin2019colloquium}. The MBL has many exotic properties. For instance, due to the existence of the local integrals of motion, it avoids the fate of thermalization \cite{huse2014phenomenology}. Naturally, depending on the disorder intensity, it divides the phases of a matter into the ergodic or thermal phase (which satisfies the eigenstate thermalization hypothesis \cite{deutsch1991quantum, srednicki1994chaos}) and the MBL phase. The entanglement entropy of the MBL eigenstates obeys an area law, whereas a volume law is held for the thermal \cite{abanin2019colloquium}. The entanglement entropy shows a power law increase over time in the thermal phase but grows logarithmically in the MBL phase. This is also a specific trait of MBL that marks a difference from the Anderson localization phase, whose entanglement entropy is bounded \cite{vznidarivc2008many, bardarson2012unbounded}. There are Poisson distributions of the energy gap in MBL and the existence of a mobility edge and so on \cite{alet2018many}. The experiments for realizing the MBL have taken place on various platforms, including ultracold atoms system \cite{schreiber2015observation}, ultracold ions \cite{smith2016many}, and superconducting circuits system \cite{roushan2017spectroscopic, xu2018emulating}. The current studies concentrates mostly on the system isolated from the environment. Only a few studies look at how the environment's dissipative effect affects the MBL/AL \cite{levi2016robustness,fischer2016dynamics,medvedyeva2016influence,vakulchyk2018signatures}. Under the framework of the Lindblad equation, they found that dissipation eventually destroys localization, which confirms intuition, and that the steady state density operator is the normalized identity. But on the way to this state, systems with MBL and non-MBL Hamiltonians behave notably differently \cite{vakulchyk2018signatures}.  

Beyond the weak coupling and Born-Markov approximation, the AA model strongly coupled with a bath is investigated. The ends of the chain are maximally entangled, whereas the remaining parts are at N\'{e}el state initially. The dynamics of imbalance and entanglement of the ends are calculated, where the imbalance is a good indicator to identify the phases of matter. Limited to the computation source, only finite time dynamics can be explored. When the AA chain is isolated from the environment, the  temporal average of imbalance is very close to zero in the ergodic phase, but that in MBL/AL retains a finite value. Meanwhile, the entanglement behaves quite differently in the ergodic and MBL/AL phases. Rapid decay to zero, resuscitation and subsequent death of the concurrence occurs in the ergodic phase. Contrarily, in the MBL/AL phase, the concurrence fluctuates around a certain value. Once there is a bath coupled with the first site of the chain, the imbalance decays to zero more rapidly in the ergodic phase. However, it is only at relatively high temperatures that the environment significantly changes the imbalance dynamics in localization phases. That is to say, the imbalance in the early evolution can still be an effective observable to detect localization or the ergodic phases. We explain it in terms of whether the eigenmodes of the system can resonate with the environment and lead to energy exchange. The entanglement of the ends will disappear quickly due to the direct interaction between the ends of the chain and bath. Additionally, when the environment is not directly coupled to the entangled ends but to intermediate sites, the entanglement of the chain's ends endures for a considerable amount of time, which inspires us to exploit the disordered environment as a buffer to safeguard quantum correlations.

This paper is organized as follows. In Section~\ref{Model and the Evolution}, we introduce our model. We then derive the influence functional and illustrate it in the tensor network language. We focus on the PT-TEBD algorithm to implement the system evolution. In Section~\ref{Results}, we introduce some quantum correlations and study them in different scenarios. In Section~\ref{open MBL}, we investigate the dissipative dynamics of imbalance and entanglement of the AA chain. Finally, we draw our conclusions in Section~\ref{Conclusion}.

\section{The tensor network method for studying the non-Markovian dynamics of a two-qubit system in an environment: Model and Evolution\label{Model and the Evolution}}

\begin{figure}[!ht]
    \centering
\includegraphics[width=6in]{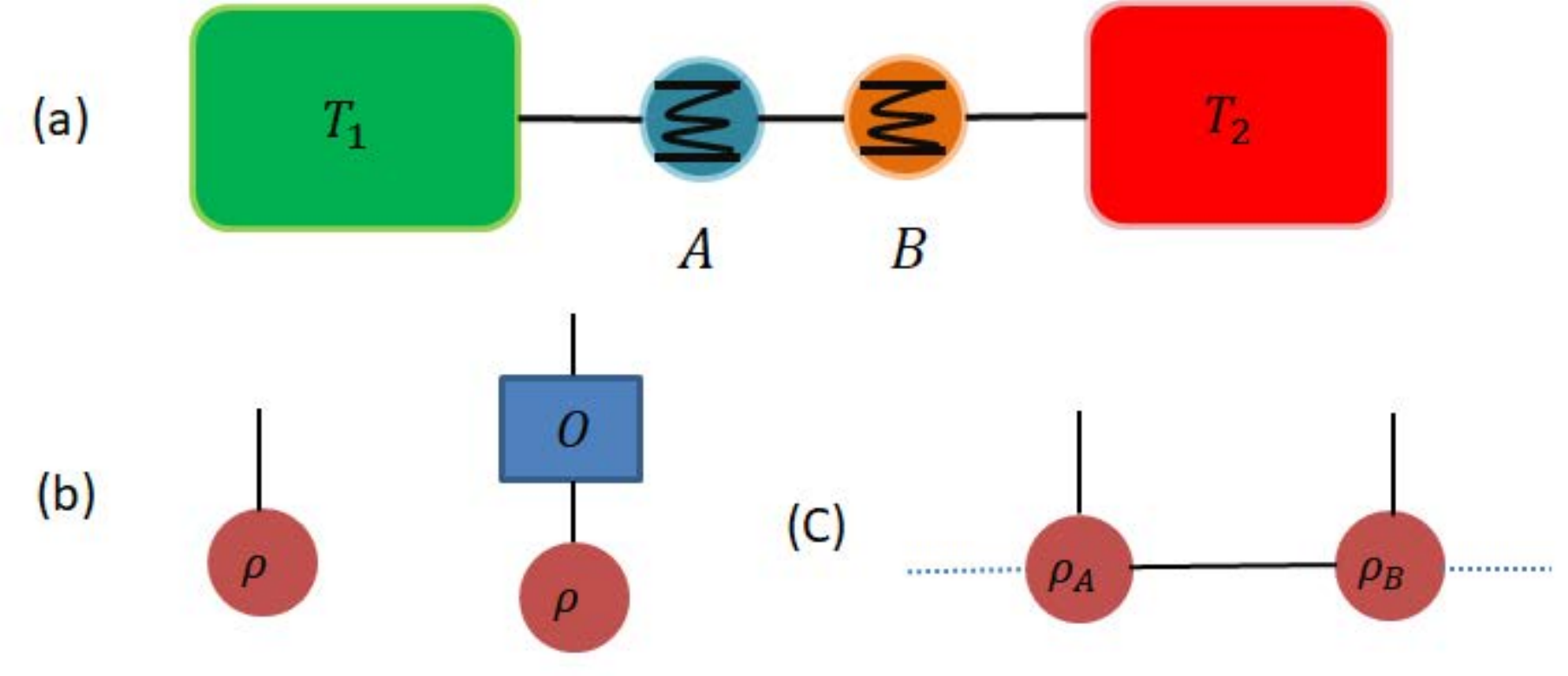}
\caption{\label{fig:1}(a) A sketch of the model; (b) Left: a rank-one tensor or vector; Right: a second-rank tensor contracts with a vector and the connecting leg represents a contraction in the tensor network language; (c) The augmented matrix product state (MPS) for the two-qubit system.}
\end{figure}

Let's begin with a simple model, which consists of two interacting qubits coupled to corresponding baths, as shown in Fig. \ref{fig:1}(a). The total Hamiltonian is given as
\begin{equation}
\label{eq:1}
H_{total}=H_{S}+H_{SB}=H_{S}+H_{B}+H_{I}.
\end{equation}
The Hamiltonian of the system is
\begin{equation}
\label{eq:2}
H_{S}=\frac{\omega_{1}}{2}\sigma_{z}^{1}+\frac{\omega_{2}}{2}\sigma_{z}^{2}+J\sigma_{x}^{1}\sigma_{x}^{2},
\end{equation}
where $\sigma_{i}^{1}=\sigma_{i}\otimes \textbf{I}$ and $\sigma_{i}^{2}=\textbf{I}\otimes\sigma_{i}$, and $\sigma_{i}$ are the Pauli matrices. The upper index `1' (`2') stands for the A(B) qubit. The energy gaps of the qubits are given as $\omega_{1}=\omega_{2}=1$, and the coupling $J=0.375$ measures the strength of the inter-qubit interaction.
The Hamiltonian of the environment plus the interaction between a single qubit and the corresponding interacting bath are given as
\begin{equation}
\label{eq:3}
H_{B}^{\alpha}+H_{I}^{\alpha}=\sum_{k}\omega_{k}^{\alpha}\hat{a}_{k}^{\alpha\dagger}\hat{a}_{k}^{\alpha}+\sigma_{x}^{\alpha}\sum_{k}(g_{k}^{\alpha}\hat{a}_{k}^{\alpha\dagger}+g_{k}^{\alpha*}\hat{a}_{k}^{\alpha}).
\end{equation}

$g_{k}$ is the coupling constant between the qubit and the environment. $\hat{a}_{k} (\hat{a}_{k}^{\dagger})$ is the annihilation (creation) operator of the bosonic environment. The upper index $\alpha$ indicates the relevant bath. We assume that the system and baths are separable at the initial time and that the baths are initially in Gaussian states, e.g., thermal equilibrium at different temperatures $T_{\alpha}$. Meanwhile, the two qubits are also separable initially and are both prepared in the ground state. A sketch of the model is shown in Fig. \ref{fig:1}(a).

%\begin{tikzpicture}
%\draw[rounded corners,color=blue] (1,0) rectangle (4,2);
%\draw (4,1)--(4.5,1);
%\draw (5,1) circle (0.5);
%\draw (5.5,1)--(6,1);
%\draw (6.5,1) circle (0.5);
%\draw (7,1)--(7.5,1);
%\draw[rounded corners,color=red] (7.5,0) rectangle (10.5,2) ;
%\end{tikzpicture}

%\begin{tikzpicture}[node distance=20pt]
  %\node[draw, rounded corners,color=blue]                        (cold)   {$T_{1}$};
 % \node[draw, circle, right=of cold]                         (qubit 1)  {A};
%  \node[draw, circle, right=of qubit 1]                        (qubit 2)  {B};
 % \node[draw, rounded corners, color=red, right=of qubit 2]     (hot)  {$T_{2}$};
  %\draw[-] (cold)  -- (qubit 1);
%  \draw[-] (qubit 1) -- (qubit 2);
 % \draw[-] (qubit 2) -- (hot);
%\end{tikzpicture}

\iffalse
\begin{tikzpicture}
[L1Node/.style={rounded corners,   draw=blue!50, fill=blue!20, very thick, minimum width =30mm, 
minimum height =20mm},
L2Node/.style={rounded corners,draw=red!50,fill=red!20,very thick,minimum width =30mm, 
minimum height =20mm},
L3Node/.style={circle,    fill=green!20, very thick, minimum size=15mm},
L4Node/.style={circle,fill=green!20,very thick, minimum size=15mm}]
\node[L1Node] (cold) at (5, 0){$T_{1}$};
\node[L2Node] (hot) at (13, 0){$T_{2}$};
\node[L3Node] (qubit 1) at (8, 0){$A$};
\node[L4Node] (qubit 2) at (10, 0){$B$};
\draw[-] (cold)  -- (qubit 1);
 \draw[-] (qubit 1) -- (qubit 2);
 \draw[-] (qubit 2) -- (hot);
\end{tikzpicture}
\fi

We work in Liouville space in the following calculation, i.e., the super-operators act on the vectorized density matrices. This is illustrated in Fig. \ref{fig:1}(b). The solid red circle with one leg is called a rank-one tensor or vector in the tensor network language \cite{tn1}.
\begin{equation}
\label{eq:4}
\rho = \sum_{ij}\rho_{ij}\ket{i}\bra{j}\stackrel{vectorized}{\longrightarrow}
|\rho \rangle \rangle= \sum_{ij}\rho_{ij} |i\rangle\otimes|j\rangle
\end{equation}
The trace is reformulated as $Tr{\cdot}=\sum_{k}\langle \langle k,k|\cdot = \langle \langle \emph{1}|\cdot$, where $\langle \langle k,k|=\langle k|\otimes\langle k|$ and $|\mathbf{1}\rangle \rangle$ is the vectorized unity matrix.
The operator acts on the density matrix and can be reformulated as 
\begin{equation}
\label{eq:5}
\left\{
\begin{aligned}
O\rho \longrightarrow\textbf{\emph{O}}^{L}|\rho \rangle \rangle = O\otimes \textbf{I}|\rho \rangle \rangle\\
\rho O\longrightarrow\textbf{\emph{O}}^{R}|\rho \rangle \rangle = \textbf{I}\otimes O^{T}|\rho \rangle \rangle
\end{aligned}
\right\}
\end{equation}
in Liouville space. This is illustrated in Fig. \ref{fig:1}(b), where a second-rank tensor contracts with a vector, where the connecting leg indicates contraction.
The time evolution of the total system is governed by the Liouville operator $\mathscr{L}=-i[\hat{H},\cdot]$.
\begin{equation}
\label{eq:6}
\rho(t)=e^{\mathscr{L}t}\rho(0)
\end{equation}
We separate $\mathscr{L}_{total}=\mathscr{L}_{S}+\mathscr{L}_{SB}$, where $\mathscr{L}_{S}=-i[\hat{H}_{s},\cdot]$ represents the pure system part and $\mathscr{L}_{SB}$ represents the remaining part, including the interaction and environmental parts. We discretize time into $N$ uniform steps and perform second-order Suzuki--Trotter splitting \cite{Suzuki-Trotter_splitting} between the system Liouville operator and the environmental Liouville operator:
\begin{equation}\begin{split}
\label{eq:7}
e^{\mathscr{L}_{total}t}&\approx [e^{\mathscr{L}_{total}\delta t}]^N\\
&=[e^{\mathscr{L}_{S}\frac{\delta t}{2}}e^{\mathscr{L}_{SB}\delta t}e^{\mathscr{L}_{S}\frac{\delta t}{2}}]^N + \mathcal{O}(\delta t^3)
\end{split}
\end{equation}
The basis of the whole system is spanned by the product space $|{s_{i}},{b_{i}}\rangle \rangle=|s_{i}^{1},s_{i}^{2},b_{i}^{1},b_{i}^{2}\rangle \rangle$ at the $i$th time step.
 The initial state can be represented by $|\rho_{total}(s_{0},b_{0})\rangle \rangle=|\rho_{s}^{1}(s_{0}^{1})\rho_{s}^{2}(s_{0}^{2})\rangle \rangle\otimes|\rho_{b}^{1}(b_{0}^{1})\rho_{b}^{2}(b_{0}^{2})\rangle \rangle$.
We trace out the degree of freedom of the baths on both sides of Eq. \ref{eq:6}, and then insert $\sum_{i}|s_{i},b_{i}\rangle \rangle\langle \langle s_{i},b_{i}|$ at the $i$th time step:

\begin{equation}\begin{split}
\label{eq:8}
\rho_s(s_N')=& Tr_{b_N'} \biggl[ \sum_{s_{N}} \sum_{b_{N}}\sum_{s_{N-1}} \sum_{b_{N-1}}  \dots \sum_{s_{0}} \sum_{b_{0}} 
\langle \langle s_N'|  e^{\mathscr{L}_s \frac{\delta t}{2}} |s_{N},b_{N} \rangle \rangle
\langle \langle s_{N},b_{N}| e^{\mathscr{L}_{bs}\delta t} e^{\mathscr{L}_s\delta t}   \\
&|s_{N-1},b_{N-1} \rangle \rangle \times \dots\langle \langle s_1,b_1| e^{\mathscr{L}_{bs}\delta t} e^{\mathscr{L}_s\frac{\delta t}{2}}
|s_{0},b_{0} \rangle \rangle \times\langle \langle s_0,b_0|\rho(s_{0},b_{0}) \rangle \rangle \biggr]\\
=&Tr_{b_N'} \biggl[ \sum_{s_{N},b_{N}} \dots \sum_{s_{0},b_{0}}\sum_{s_{N-1}'}  \dots \sum_{s_{0}'} 
\langle \langle s_N'|  e^{\mathscr{L}_s \frac{\delta t}{2}}|s_{N},b_{N} \rangle \rangle\langle \langle s_{N},b_{N}| e^{\mathscr{L}_{bs}\delta t} |s_{N-1}' \rangle \rangle\\
&\langle \langle s_{N-1}'|e^{\mathscr{L}_{s}\delta t}| s_{N-1},b_{N-1}\rangle \rangle \times \dots
\langle \langle s_1,b_1| e^{\mathscr{L}_{bs}\delta t} |s_{0}'\rangle \rangle \langle \langle s_0'|e^{\mathscr{L}_s\frac{\delta t}{2}}
|s_{0},b_{0} \rangle \rangle \times\langle \langle s_0,b_0|\rho(s_{0},b_{0}) \rangle \rangle \biggr]\\
=&\sum_{s_{N},b_{N}} \dots \sum_{s_{0},b_{0}}\sum_{s_{N-1}'}  \dots \sum_{s_{0}'}  \langle \langle s_N'|  
e^{\mathscr{L}_s \frac{\delta t}{2}}|s_{N} \rangle \rangle 
\langle \langle s_{N-1}^{'}|e^{\mathscr{L}_s\delta t}  |s_{N-1} \rangle \rangle
\times \dots
\langle \langle s_{0}^{'}|e^{\mathscr{L}_s\frac{\delta t}{2}}
|s_{0}\rangle \rangle \\
&Tr_{b_N'} \biggl[\langle \langle s_{N},b_{N}| e^{\mathscr{L}_{bs}\delta t} |s_{N-1}',b_{N-1} \rangle \rangle  \langle \langle s_{N-1},b_{N-1}| e^{\mathscr{L}_{bs}\delta t} |s_{N-2}',b_{N-2} \rangle \rangle \dots \\
&\langle \langle s_1,b_1| e^{\mathscr{L}_{bs}\delta t}|s_{0}^{'},b_{0} \rangle \rangle|b_{N} \rangle \rangle\langle \langle s_0,b_0|\rho(s_{0},b_{0}) \rangle \rangle\biggr] \\
=&\sum_{s_{N},b_{N}} \dots \sum_{s_{0},b_{0}}\sum_{s_{N-1}'}  \dots \sum_{s_{0}'} \langle \langle s_N'|  
e^{\mathscr{L}_s \frac{\delta t}{2}}|s_{N} \rangle \rangle 
\langle \langle s_{N-1}'|e^{\mathscr{L}_s\delta t}  |s_{N-1} \rangle \rangle
\times \dots
\langle \langle s_{0}'|e^{\mathscr{L}_s\frac{\delta t}{2}}
|s_{0}\rangle \rangle\langle \langle s_0|\rho(s_{0}) \rangle \rangle \\
&\mathscr{I}(s_{1},s_{1}^{'};\dots s_{N},s_{N}^{'})
\end{split}
\end{equation}

where $\mathscr{I}(s_{1},s_{1}^{'};\dots s_{N},s_{N}^{'})$ is the influence functional \cite{IF1}. 

\begin{equation}\begin{split}
\label{eq:9}
\mathscr{I}(s_{1},s_{1}^{'};\dots s_{N},s_{N}^{'}) =& Tr_{b_N'} \biggl[\langle \langle s_{N},b_{N}| e^{\mathscr{L}_{bs}\delta t} |s_{N-1}',b_{N-1} \rangle \rangle  \langle \langle s_{N-1},b_{N-1}| e^{\mathscr{L}_{bs}\delta t} |s_{N-2}',b_{N-2} \rangle \rangle \dots \\
&\langle \langle s_1,b_1| e^{\mathscr{L}_{bs}\delta t}|s_{0}',b_{0}\rangle \rangle |\rho(b_{0}) \rangle \rangle \biggr]
\end{split}
\end{equation}

\begin{figure}[!ht]
    \centering
\includegraphics [width=6in] {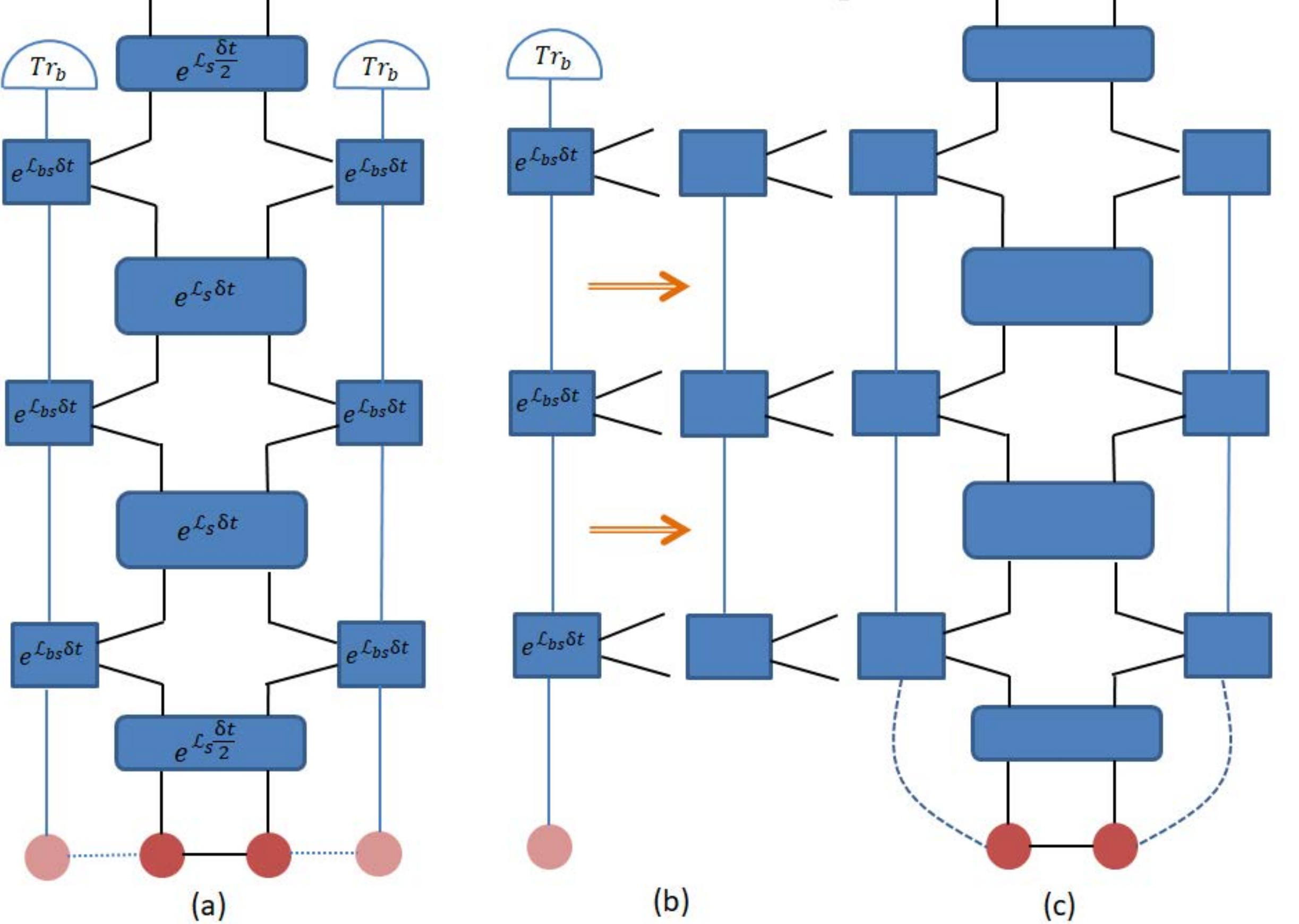}
\caption{\label{fig:2}(a) A graphic illustration of Eq. \ref{eq:8} in three time steps; (b) The influence functional contracts to the process tensor; (c) The evolution of the system with the process tensor, which is TEBD-like.}
\end{figure}

Any n-dimensional array, such as the density matrix in Fig. \ref{fig:1}(b), can be called an n-rank tensor in tensor network language. Its expansion coefficient under an orthonormal basis can be considered a tensor for a general quantum state. We can utilize tensor train decomposition to obtain its MPS form \cite{tn1,tn2}. With the help of the augmented matrix product state (aMPS \cite{thermalization,pt2,pt3}) in Fig. \ref{fig:1}(c), we formally construct an MPS for the two-qubit system, where the augmented legs measure the correlation between the system and the two baths. In our case, the bond dimensions of the augmented legs are equal to one, which means that there is no initial correlation between the system and the two baths. Equation \ref{eq:8} is graphically illustrated in Fig. \ref{fig:2}(a). The network is composed of the total density matrix, which is in the MPS form (the qubits and baths are represented by respective sites), the pure system Liouville tensor $\langle \langle s_i'|  
e^{\mathscr{L}_s \frac{\delta t}{2}}|s_{i} \rangle \rangle$, and the environmental Liouville tensor $\langle \langle s_{i},b_{i}| e^{\mathscr{L}_{bs}\delta t} |s_{i-1}',b_{i-1} \rangle \rangle$. The pure system Liouville tensor contains the interaction between the two qubits. Therefore, it has four legs indexed by $s_i^{1'},s_i^{2'},s_i^{1}$, and $s_i^{2}$. There is no interaction between the two baths, so their action can be separated. The environmental Liouville tensor for the single bath also has four legs indexed by $s_{i-1}^{\alpha'},s_i^{\alpha},b_{i-1}^{\alpha}$, and $b_i^{\alpha}$. The solid blue line in Fig. \ref{fig:2}(a) represents the degree of freedom of the baths $b_{i}$, and these legs should only connect to themselves. The solid black line in Fig. \ref{fig:2}(a) represents the degree of freedom of the qubits $s_{i}$. We connect the same indices from the MPS of the system according to Eq. \ref{eq:8}. The same indices being connected indicates summation. Finally, we trace the degree of freedom of the baths, where the trace cap is described by the semi-circle in Fig. \ref{fig:2}(a). In addition, the contraction of the influence functional produces the process tensor, as shown in Fig. \ref{fig:2}(b), which is a multi-linear map from the set of all possible control operation sequences in the lab on the system to the resulting output states \cite{pt1}. The process tensor includes all the influences of the environment. According to the time order, we connect it to the pure system Liouville tensor. The process tensor can be recycled. Finally, the evolution of the system can be carried out with process tensor-time evolving block decimation (PT-TEBD) \cite{TEMPO3,thermalization,TEBD}, as shown in Fig. \ref{fig:2}(c). We contract the network layer by layer from the bottom to the top.

The influence functional and the resulting process tensor can both be constructed in a matrix product operator (MPO) form \cite{TEMPO1,TEMPO2,pt2,pt3}. When we choose the eigenbasis of the environmental part as the computational basis, the influence functional for a single bath can be written as

\begin{equation}\begin{split}
\label{eq:10}
\mathscr{I}(s_{1},s_{1}^{'};
\dots s_{N},s_{N}^{'})=&\exp{(-\sum_{k=1}^{N}\sum_{k'=1}^{k}(s^{+}_{k}-s^{-}_{k})(\eta_{kk'}s^{+}_{k}-\eta_{kk'}^{*}s^{-}_{k}))}\\
=&\prod_{k=1}^{N}\it{I}_{0}(s^{\pm}_{k})\prod_{k=1}^{N-1}\it{I}_{1}(s^{\pm}_{k+1},s^{\pm}_{k})\dots\prod_{k=1}^{1}\it{I}_{N}(s^{\pm}_{k-1+N},s^{\pm}_{k})
\end{split}
\end{equation}
where $\it{I}_{m}=\exp{(-(s^{+}_{k+m}-s^{-}_{k+m})(\eta_{k+m,k}s^{+}_{k}-\eta_{k+m,k}^{*}s^{-}_{k}))}$ and $|s^{\pm}_{k}\rangle$ satisfies $\sigma_{x}|s^{\pm}_{k}\rangle=s^{\pm}_{k}|s^{\pm}_{k}\rangle$ at time $t_{k}$. The detailed expressions for $\eta_{kk'}$ can be found in \cite{IF2,IF3}. $\eta_{kk'}$ is dependent on the spectrum density $J(\omega)$:
\begin{equation}\begin{split}
\label{eq:11}
J(\omega)=2\alpha\frac{\omega^{\zeta}}{\omega_{c}^{\zeta-1}}e^{-\frac{\omega}{\omega_{c}}},
\end{split}
\end{equation}
where $\omega_{c}$ is the frequency cutoff. The reservoirs are Ohmic when $\zeta=1$, sub-Ohmic when $\zeta<1$, and super-Ohmic when $\zeta>1$. Under the sub-Ohmic spectrum, the lower frequencies $\omega<\omega_{c}$ dominate. Under the super-Ohmic spectrum, the higher frequencies $\omega>\omega_{c}$ dominate. The low-frequency behavior is described by $J(\omega)\sim \omega^{\zeta}$. 

\begin{figure}[!ht]
    \centering
\includegraphics [width=6in] {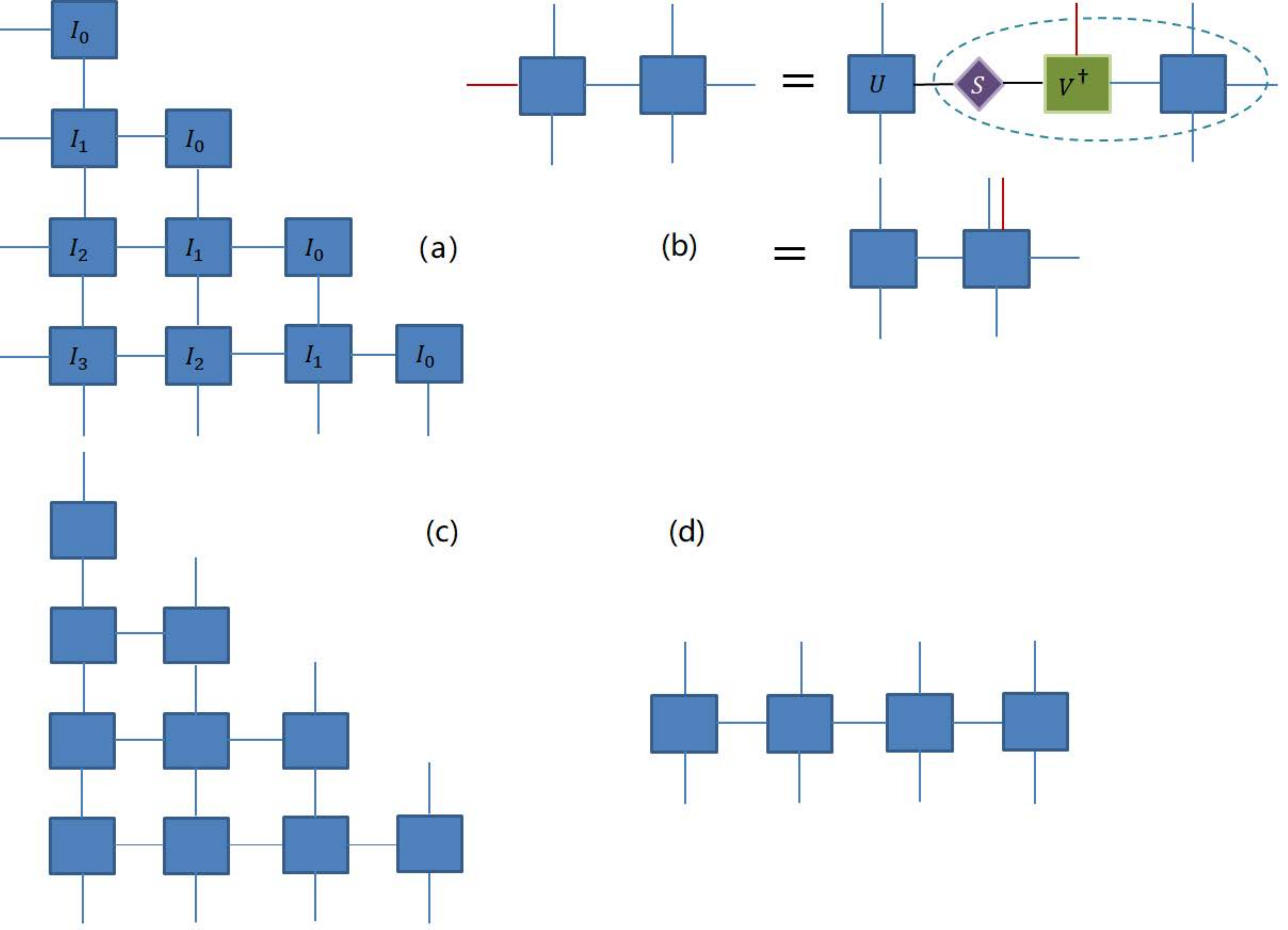}
\caption{\label{fig:3} (a) A graphic illustration of Eq. \ref{eq:10} when $N=3$; (b) SVD decomposition and contraction; (c) By repetitive SVD decomposition and contraction, we can reshape the tensor in (a) into this one; (d) The final desired MPO form of the process tensor.}
\end{figure}

Equation \ref{eq:10} can be translated into an MPO form, where $\it{I}_{0}$ is represented by a second-rank tensor and $\it{I}_{m>0}$ is represented by a fourth-rank tensor, as shown in Fig. \ref{fig:3}(a). The legs representing the same time points are connected. Through repetitive SVD decomposition and contraction, as shown in Fig. \ref{fig:3}(b), we can reshape the tensor in Fig. \ref{fig:3}(a) into the one in Fig. \ref{fig:3}(c). Finally, we contract the tensor layer by layer from the bottom to the top in Fig. \ref{fig:3}(c) to derive the desired MPO-form tensor in Fig. \ref{fig:3}(d). In practice, one usually implements a memory cutoff to save on computational costs. The cutoff for a specific bath depends on its correlation properties. The memory cutoff $\tau_{c}$ dictates how long the system histories are kept to capture the non-Markovianness. We choose $\it{I}_{m}= \textbf{I}\otimes\textbf{I}$ once $m\delta t$ exceeds $\tau_{c}$.

\begin{figure}[!ht]
    \centering
\includegraphics [width=2in] {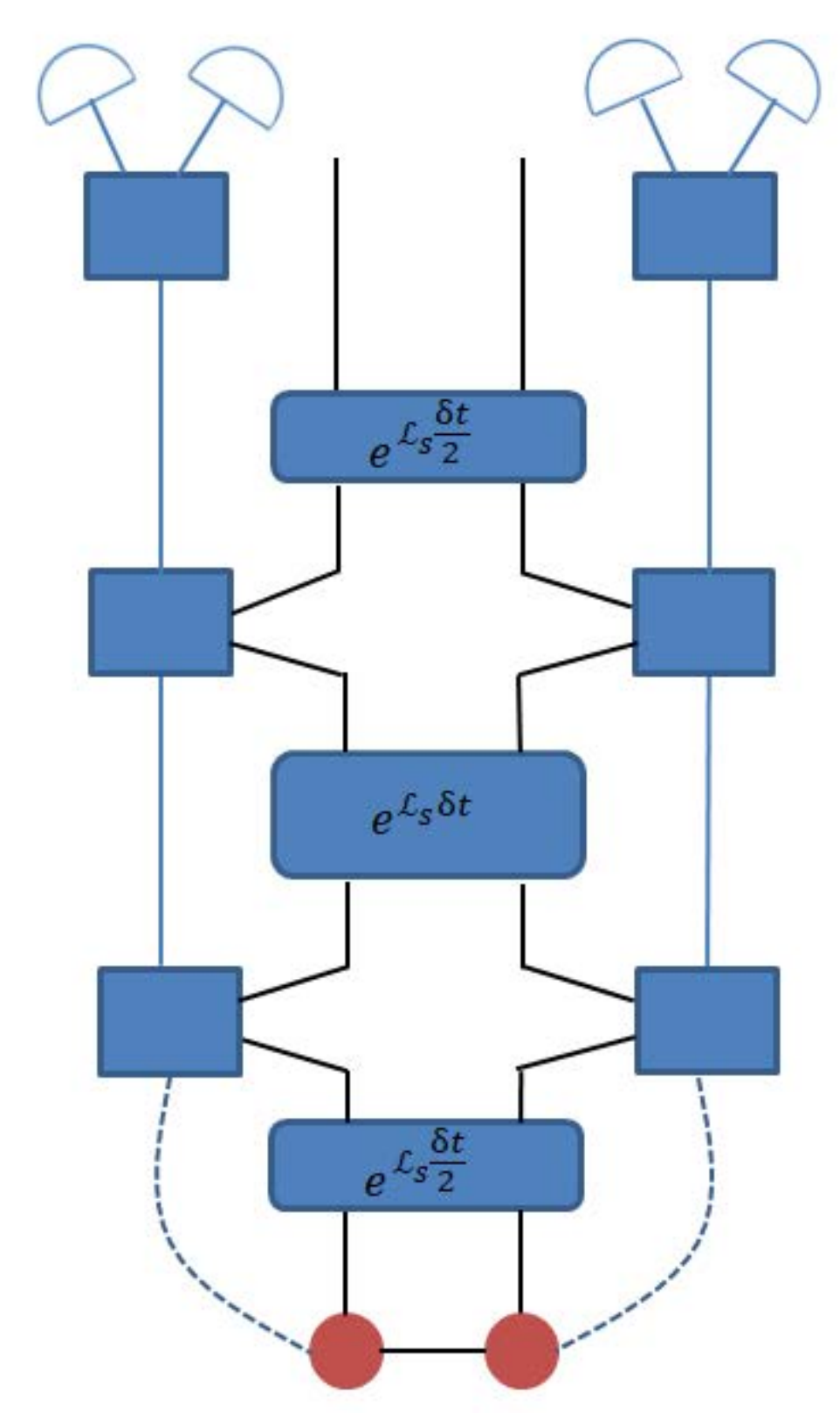}
\caption{\label{fig:4} A graphic illustration of how to derive the intermediate time evolution at the second time step when the total time steps $N=3$. The trace cap is $|\mathbf{1}\rangle \rangle$.}
\end{figure}

There are three error sources when we run the PT-TEBD program. The first originates from the second-order Suzuki--Trotter split, which causes the third- and higher-order error $\mathcal{O}(\delta t^3)$. The second type of error comes from the low-rank matrix approximation when we derive the process tensor and perform time-evolving block decimation. Suppose $X=USV^{\dagger}$, which is the exact SVD decomposition. We throw away some lower singular values in $S$ and corresponding vectors in $U$ and $V$. The new $\tilde{X} =\tilde{U}\tilde{S}\tilde{V}^{\dagger}$, and the associated error is $||\tilde{X}-X||_{2}<\epsilon\max{S}$. The third type of error is associated with the memory cutoff. We throw away small time correlations to save computational resources. In the PT-TEBD procedure, we first construct a process tensor for the baths. The PT can then be recycled for the different initial states of the system. We then apply the TEBD algorithm to realize the time evolution. So far, we can only derive the result at the $N$th time step, but we are also interested in the intermediate time evolution. In this situation, we can trace out the latter time legs to obtain the information at intermediate times, as shown in Fig. \ref{fig:4}. This means we use the trace cap on the legs of the system after the desired time step. This approach is permitted by the containment property of the process tensor, which says that if $k \geq k_0 \geq j_0 \geq j$, the process tensor
$T_{k_0:j_0}$ is contained in $T_{k:j}$ \cite{pt1}. %More information about the PT-TEBD algorithm can be found in \cite{thermalization,TEMPO3}

The main features of the PT-TEBD algorithm are that it is numerically exact and can be applied over wide ranges as long as the environment is a thermal equilibrium bosonic bath. It can also dispose of the time-dependent system \cite{optimal_control} and the one-dimensional chain system \cite{thermalization} beyond the simple single-qubit system. Starting from the QUAPI approach and without any other approximations, the propagation of the ADT that encodes the system’s history replaces that of the density matrix. The ADT grows at each time step, so we use the finite memory approximation to limit its size \cite{IF2,IF3}. The number of elements in the ADT scales exponentially with the memory cutoff. If the full tensor is considered, one quickly encounters memory problems. Fortunately, an MPS is a natural tool with which to represent a high-rank tensor efficiently. The memory required to store a tensor scales exponentially with its rank but scales as a polynomial in the MPS representation. The original TEMPO algorithm describes the propagation of the ADT in MPS form \cite{TEMPO1}. The subsequent PT-TEBD algorithm inherits this highly efficient characteristic and integrates the IF as a process tensor in MPO form \cite{TEMPO3,optimal_control,nonadditive}. Another advantage of the process tensor is that it is independent of the system Hamiltonian and the initial state of the system. In addition, it can be recycled for different system settings \cite{optimal_control}. The evolution of the system is computed by a highly efficient and well-developed TEBD algorithm.

\section{Non-Markovian quantum correlations under equilibrium and nonequilibrium environments\label{Results}}
\subsection{Measures of the quantum correlations}

In this subsection, we introduce certain important measures of the quantum correlations. Coherence, being at the heart of interference phenomena, plays a cornerstone role in quantum physics as it enables applications that are impossible within classical mechanics. It can be measured as~\cite{Quantifying_Coherence}
\begin{equation}
\label{eq:12}
\mathscr{C}_{l_{1}}=\sum_{i\neq j}\mid\rho_{ij}\mid.
\end{equation}

Quantum entanglement is a major resource for accomplishing quantum information processing tasks such as teleportation \cite{Teleporting}, quantum key distribution \cite{Quantum_Cryptography}, and quantum computing \cite{Quantum_entanglement}. Among the many measures of entanglement of a two-qubit system, the concurrence is extensively used in many contexts. The concurrence of a two-qubit mixed state $\rho$ is defined as \cite{Entanglement_of_a_Pair_of_Quantum_Bits}
\begin{equation}
\label{eq:13}
\mathscr{C}=Max(0,\lambda_{1}-\lambda_{2}-\lambda_{3}-\lambda_{4}),
\end{equation}

where $\lambda_{i}$ represents the square root of the $i$th eigenvalue, in descending order of the matrix $\rho\widetilde{\rho}$, with $\widetilde{\rho}=(\sigma_{2}\bigotimes\sigma_{2})\rho^{T}(\sigma_{2}\bigotimes\sigma_{2})$, while $T$ denotes transposition.

Another important quantum correlation measure is the quantum discord~\cite{Quantum_Discord}. This measures the non-classical correlation between two subsystems of a quantum system. The discord includes correlations due to quantum physical effects but does not necessarily involve the concept of quantum entanglement. In fact, it is a different type of quantum correlation to the entanglement because separable mixed states (that is, with no entanglement) can have a non-zero quantum discord. The quantum discord is sometimes also identified as a measure of the quantumness of correlation functions. The geometric discord of a bipartite quantum state is defined as~\cite{GD}

\begin{equation}
\label{eq:14}
\mathscr{D}(\rho)=\min{||\rho-\rho_{0}||^2}_{\rho_{0}\in\Omega}
\end{equation}
where $\Omega$ denotes the set of zero-discord states and $||X-Y||^2=Tr(X-Y)^{2}$ is the square norm in Hilbert--Schmidt space. This can be evaluated for an arbitrary two-qubit state. For any two-qubit state, the density matrix is given by the following expression:

\begin{equation}
\label{eq:15}
\rho_{AB}=\frac{1}{4}(I_{a}\bigotimes I_{b}+\sum_{i=1}^{3}(a_{i}\sigma_{i}\bigotimes I_{b}+I_{a}\bigotimes b_{i}\sigma_{i})+\sum_{i,j=1}^{3}C_{ij}\sigma_{i}\bigotimes\sigma_{j}).
\end{equation}
The geometric discord is given as~\cite{GD}
\begin{equation}
\label{eq:16}
\mathscr{D}(\rho)=\frac{1}{4}(||a||^2+||C||^2)-\lambda_{max}),
\end{equation}
where $\lambda_{max}$ is the maximum eigenvalue of $aa^{T}+CC^{T}$. $a$ is the vector composed of $a_{i}$ and $C$ is the matrix composed of $C_{ij}$.

%In addition to all the above measures of quantum correlations, one could also attempt to quantify them in terms of an application, e.g., the fidelity of teleportation. Quantum teleportation is a technique that takes advantage of quantum entanglement as a teleportation channel to teleport quantum states from a sender to a receiver without transmitting qubits \cite{Teleporting}. %The standard protocol of teleportation are summarized as follows \cite{Teleporting}: (i) The first step is to establish the initial entanglement between Alice and Bob (hold qubit-2 and qubit-3 respectively) when they are together. Alternatively, it can be distributed by Charlie, a third party that produces entangled particles pairs. (ii) Alice interacts qubit-1 with her half of the entangled pair (qubit-2), and then performs a special measurement called the Bell state measurement on them. Alice sends the result of the measurement to Bob by means of a classical channel. (iii) To complete the teleportation, Bob either does nothing or performs a recovery operation (depending on Alice's measurement) on his half of the entangled pair (qubit-3) to restore the state that Alice sent. The fidelity is 1 for the ideal teleportation process, which means that the initial state is the maximal entangled state and can also maintain it at all time. We assume that the classical channel is noiseless. The entangled qubits pair interact the environment inevitably, which leads to the entanglement within system scrambles to the environment. 

\subsection{Ohmic reservoirs}

\begin{figure}[!ht]
\centering
\includegraphics[width=1\textwidth]{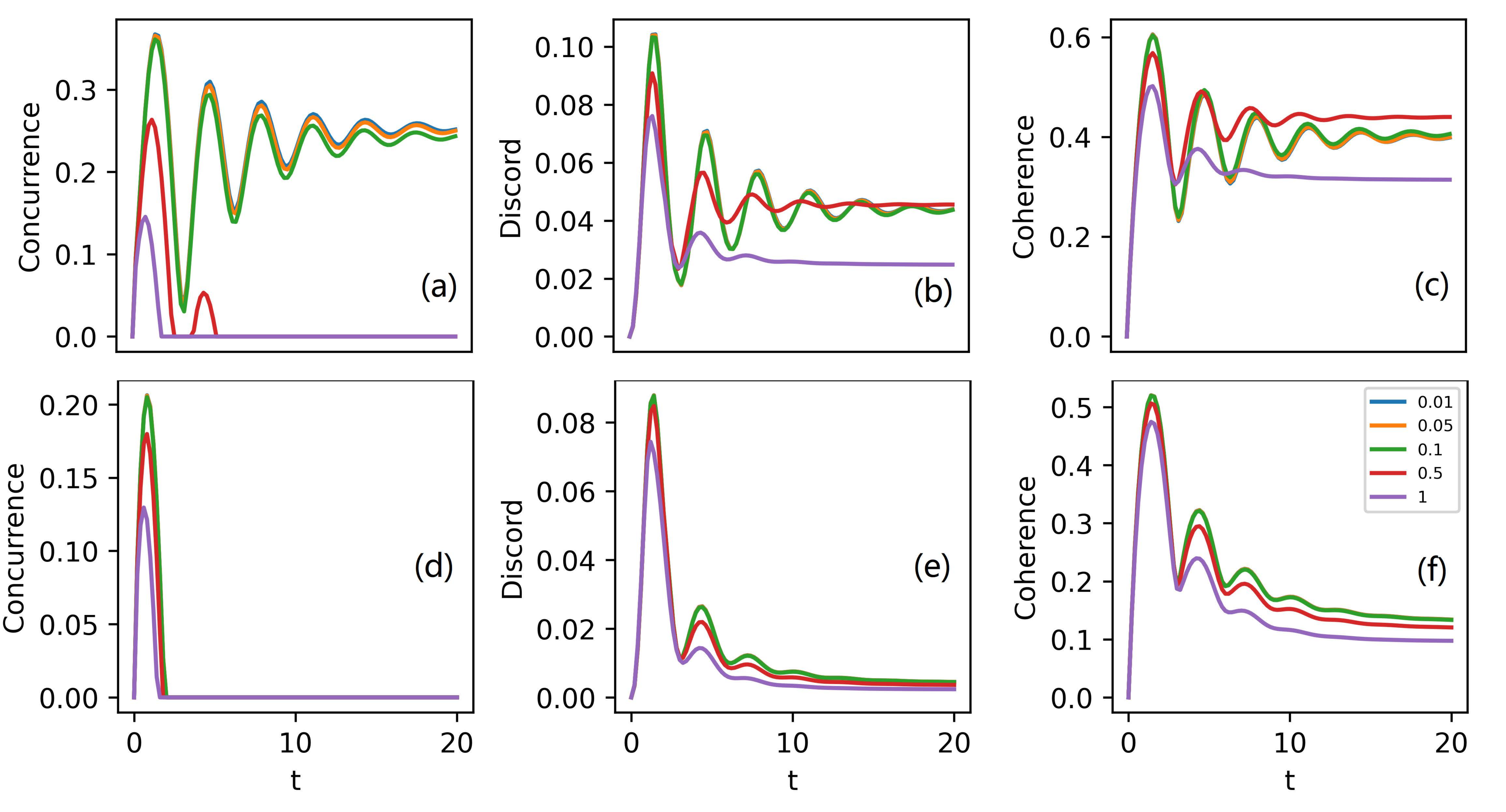}
\caption{\label{fig:O_E} The equilibrium quantum correlations in the Ohmic reservoirs at different temperatures (a--c) for non-Markovian evolution and (d--f) Markovian evolution. The other parameters are $\delta t=0.2$ for both the non-Markovian evolution and the Markovian evolution, $\zeta=1$, $\alpha=0.1$, and $\omega_c=4$. The blue and yellow solid lines overlap in (b,c). The blue, yellow, and green solid lines overlap in (d--f). }
\end{figure}

In this subsection, we study the quantum correlations in the Ohmic reservoirs under different temperatures. The memory cutoff $\delta k_{max}$ is $40$ steps, and the time interval is $\delta t=0.2$ for each step. These settings produce sufficient non-Markovian effects. The truncation error for deriving the process tensor is $\xi=10^{-5}$, and $\epsilon=10^{-6}$ for performing TEBD in this study. In Fig. \ref{fig:O_E}(a--c), we plot the quantum correlations in the Ohmic baths under different temperatures. The correlations oscillate over time, and their amplitudes decay until the correlations reach certain constants. Such oscillation is widespread in non-Markovian environments \cite{Two_Accelerated_Detectors,Entanglement_oscillations_in_non-Markovian_quantum_channels}. There are no other new phenomena, even though we prolong the evolution time. The concurrence vanishes in the higher temperature zone (e.g., $T\geq1$) but survives forever in the low-temperature regime. The entanglement suddenly dies and then reappears for a higher temperature (e.g., $T=0.5$). The sudden death and rebirth of entanglement has been observed in several different physical models (see, for instance, Refs. \cite{Non-Markovian_entanglement_preservation,Sudden_death_and_sudden_birth, Dynamics_of_interacting_qubits_coupled_to_a_common_bath}). The geometry discord and the coherence can survive at even higher temperatures. Furthermore, the geometry discord and the coherence vary non-monotonically with the temperature at certain times. We also plot the quantum correlations under the Markovian approximation (meaning that we only keep one-step memory in the evolution, i.e., $\delta k_{max}=1$ --- we comment more about this approximation in Sec. \ref{BR}) as a contrast. In Fig. \ref{fig:O_E}(d--f), the oscillations are weaker than in the non-Markovian case. The most harvesting quantum correlations under the memoryless approximation are less than those without the approximation. Therefore, memory can boost and maintain the quantum correlations under certain conditions. 

\begin{figure}[!ht]
\centering
\includegraphics[width=1\textwidth]{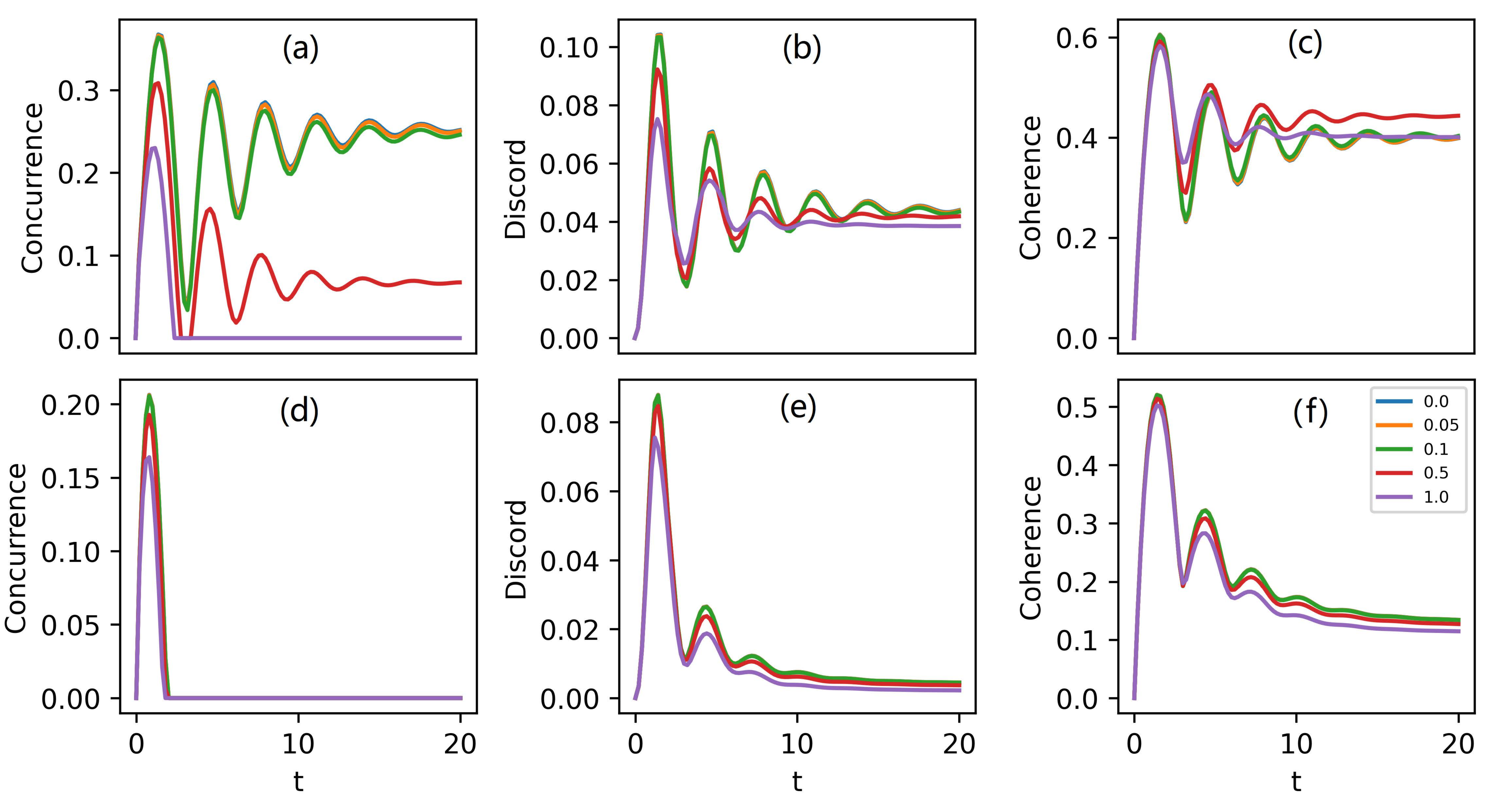}
\caption{\label{fig:O_NE}The nonequilibrium quantum correlations in the Ohmic reservoirs at different temperatures (a--c) for non-Markovian evolution and (d--f) Markovian evolution. The temperature of one of the two baths is $0.01$, and the other bath increases above $0.01$. The other parameters are $\delta t=0.2$ for the non-Markovian evolution, $\delta t=0.05$ for the Markovian evolution, $\zeta=1$, $\alpha=0.32$, and $\omega_c=4$. The blue, yellow, and green solid lines almost overlap in (a--f).}
\end{figure}

We also plot the quantum correlations under nonequilibrium in Fig. \ref{fig:O_NE}. The temperature of one of the two baths is $0.01$, while the temperature of the other bath increases above $0.01$. Compared to the equilibrium case, the trends are similar. The geometry discord and the coherence vary non-monotonically with the temperature differences at certain times in Fig. \ref{fig:O_NE}(b,c). The quantum correlations change almost monotonically with the temperature differences in Fig. \ref{fig:O_NE}(d,e,f) in the Markovian case.

\subsection{Sub-Ohmic reservoirs}
\begin{figure}[H]
\centering
\includegraphics[width=1\textwidth]{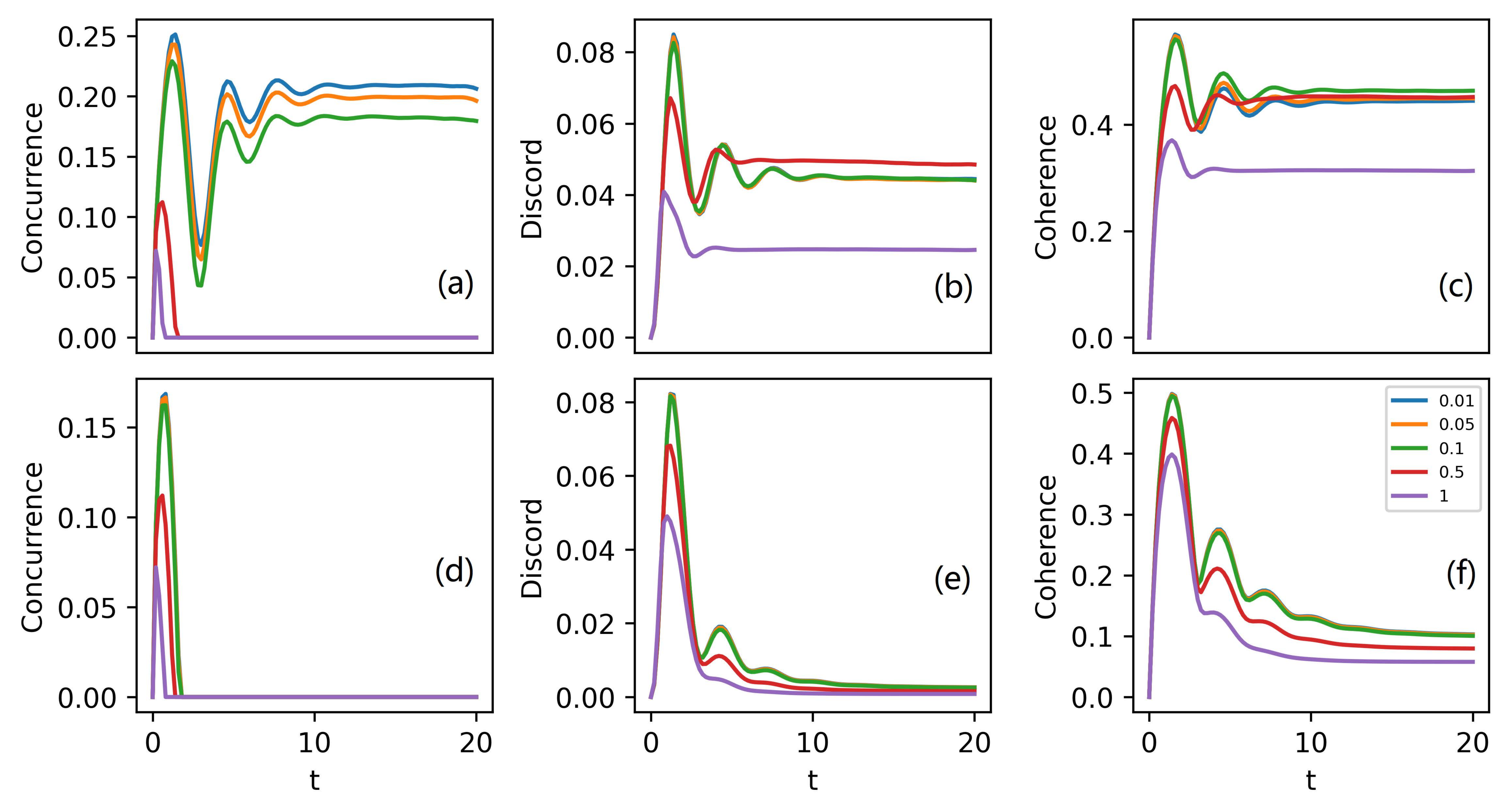}
\caption{\label{fig:sub_O_E} Equilibrium quantum correlations in the sub-Ohmic reservoirs with different temperatures (a--c) for non-Markovian evolution and (d--f) for Markovian evolution. The other parameters are $\delta t=0.2$ for both the non-Markovian evolution and the Markovian evolution, $\zeta=0.6$, $\alpha=0.1$, and $\omega_c=4$. The blue, yellow, and green solid lines overlap in (d--f).}
\end{figure}

In this subsection, we study the quantum correlations in the sub-Ohmic reservoirs with different temperatures. The memory cutoff $\delta k_{max}$ is $50$ steps, and the time interval is $\delta t=0.2$ for each step. In contrast to the Ohmic case, the quantum correlations in the sub-Ohmic baths oscillate with lower amplitudes and the amplitudes decay more rapidly in Fig. \ref{fig:sub_O_E}(a--c). The geometry discord and the coherence also vary non-monotonically with the temperature at certain times. Meanwhile, the harvesting quantum correlations are hard to retain when the environment loses memory, as shown in Fig. \ref{fig:sub_O_E}(d--f). %Compared to the Ohmics, the correlations dynamics are more sensitive to the temperature.

\begin{figure}[!ht]
\centering
\includegraphics[width=1\textwidth]{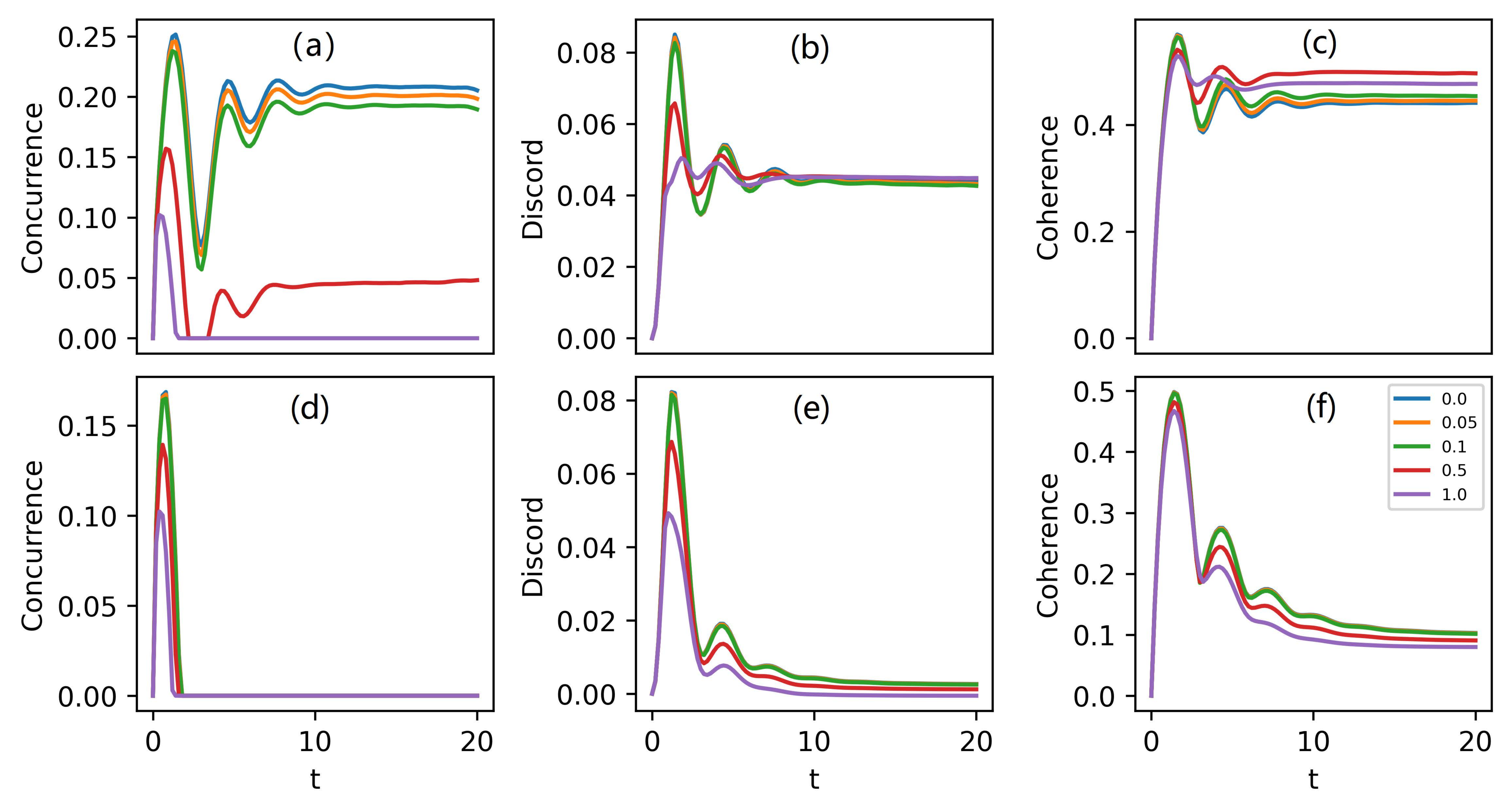}
\caption{\label{fig:sub_O_NE} Nonequilibrium quantum correlations in the sub-Ohmic baths under different temperatures (a--c) for non-Markovian evolution and (d--f) for Markovian evolution. The temperature of one of the two baths is $0.01$, and the temperature of the other bath increases above $0.01$. The other parameters are $\delta t=0.2$ for both the non-Markovian evolution and the Markovian evolution, $\zeta=0.6$, $\alpha=0.1$, and $\omega_c=4$. The blue, yellow, and green solid lines overlap in (d--f). }
\end{figure}

In the nonequilibrium scenario in Fig. \ref{fig:sub_O_NE}(a), the entanglement can still survive for a long time at a larger temperature difference $\Delta T=0.5$. In addition, the larger temperature difference can boost the correlations, as seen in Fig. \ref{fig:sub_O_NE}(b,c). However, the nonequilibrium effect is weak for the case under the memoryless approximation. The system reaches a steady state faster when the baths are 
at higher temperatures for the Ohmic and sub-Ohmic reservoirs.

\subsection{Super-Ohmic reservoirs}
\begin{figure}[H]
\centering
\includegraphics[width=1\textwidth]{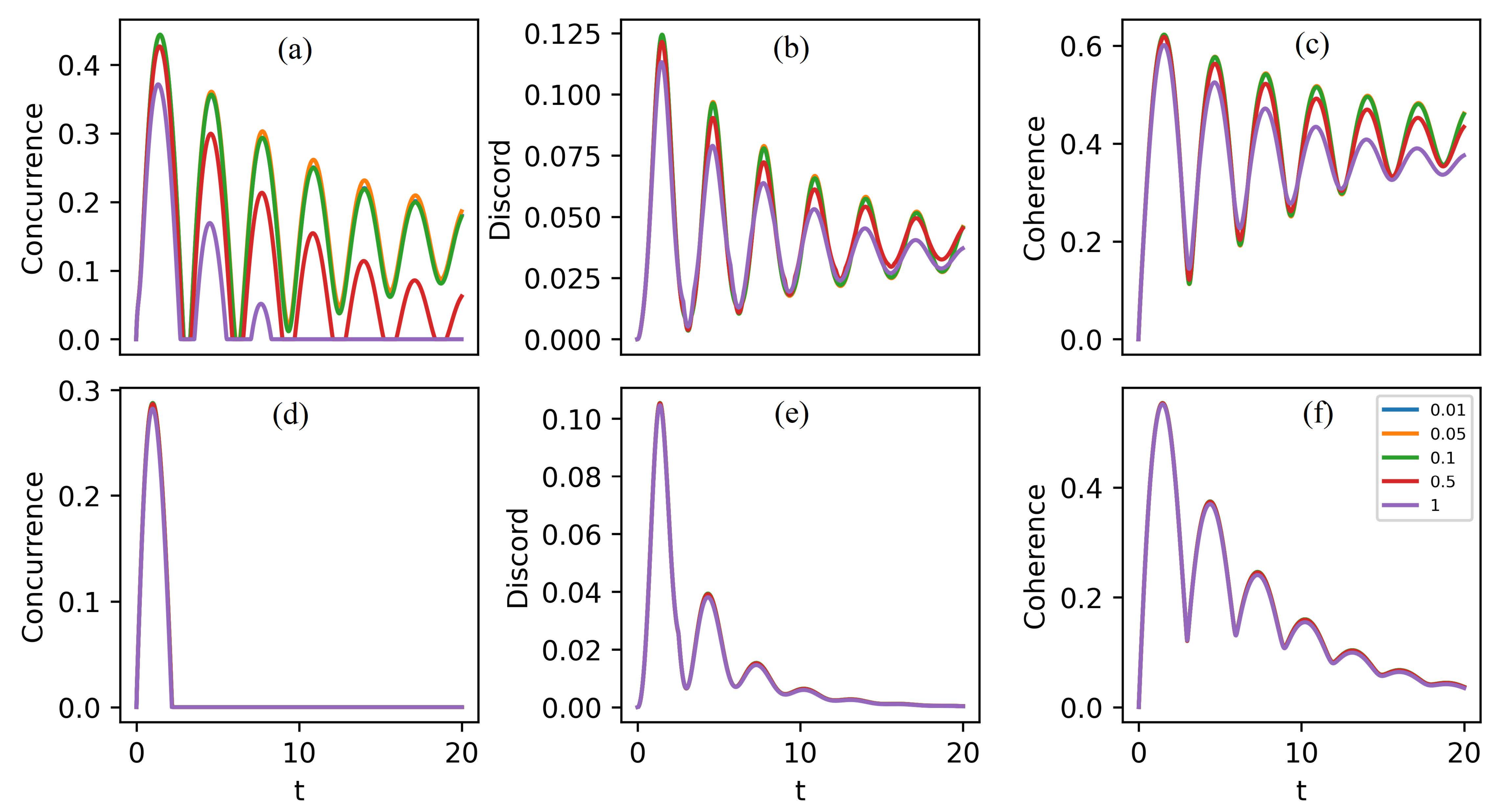}
\caption{\label{fig:super_O_E} Equilibrium quantum correlations in the super-Ohmic baths under different temperatures (a--c) for non-Markovian evolution and (d--f) Markovian evolution. The other parameters are $\delta t=0.025$ for both the non-Markovian evolution and the Markovian evolution, $\zeta=2$, $\alpha=0.1,$ and $\omega_c=4$. The blue, yellow, and green solid lines almost overlap in (a--c). All of the colored lines overlap in (d--f).}
\end{figure}

In this subsection, we study the quantum correlations in the super-Ohmic baths under different temperatures. The memory cutoff $\delta k_{max}$ is $40$ steps, and the time interval is $\delta t=0.025$ for each step. As shown in Fig. \ref{fig:super_O_E}, the oscillations of the quantum correlations are more significant here, and they are damped more slowly than in the previous two cases. The entanglement even undergoes rebirth at high temperatures in Fig. \ref{fig:super_O_E}(a). %The geometric discord and coherence are not sensitive to the temperature compared the two previous cases in the \ref{fig:super_O_E}(b,c). 
The quantum correlations of the two-qubit system in the super-Ohmic baths can reach the most significant values among the three types of baths. Under the Markovian approximation, the quantum correlations seem to be insensitive to the temperature. The dynamics of the quantum correlations under the memoryless approximation also decay more slowly than in the previous two cases. The super-Ohmic environment has the most considerable memory effect, which is consistent with Ref. \cite{Non-Markovian_entanglement_dynamics_of_noisy_continuous-variable_quantum_channels}.

\begin{figure}[H]
\centering
\includegraphics[width=1\textwidth]{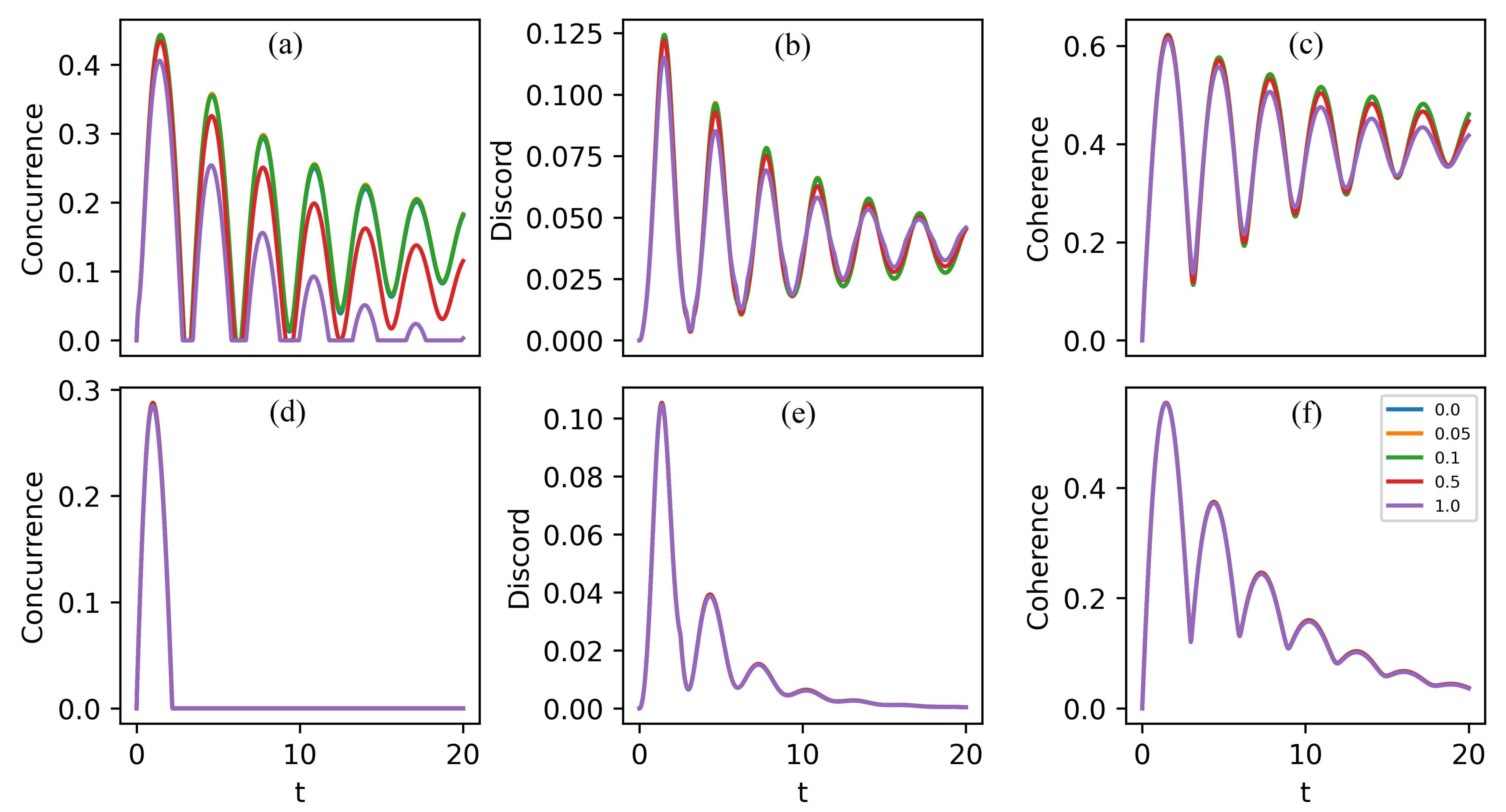}
\caption{\label{fig:super_O_NE} Nonequilibrium quantum correlations in the super-Ohmic baths under different temperatures (a--c) for non-Markovian evolution and (d--f) Markovian evolution. The temperature of one of the two baths is $0.01$, and the temperature of the other bath increases above $0.01$. The other parameters are $\delta t=0.025$ for both the non-Markovian evolution and the Markovian evolution, $\zeta=2$, $\alpha=0.1$, and $\omega_c=4$. The blue, yellow, and green solid lines almost overlap in (a--c). All of the colored lines overlap in (d--f).}
\end{figure}

As shown in Fig. \ref{fig:super_O_NE}, in the nonequilibrium scenario, the dynamics of the correlations are similar to those in the equilibrium case. The dynamics of the correlations under the memoryless approximation are insensitive to the temperature differences. We find that the reservoirs with memory maintain the quantum correlations and show a weak decoherence effect in all cases. 

\subsection{Comparison with the Bloch--Redfield master theory\label{BR}} 

\begin{figure}[H]
\centering
\includegraphics[width=1\textwidth]{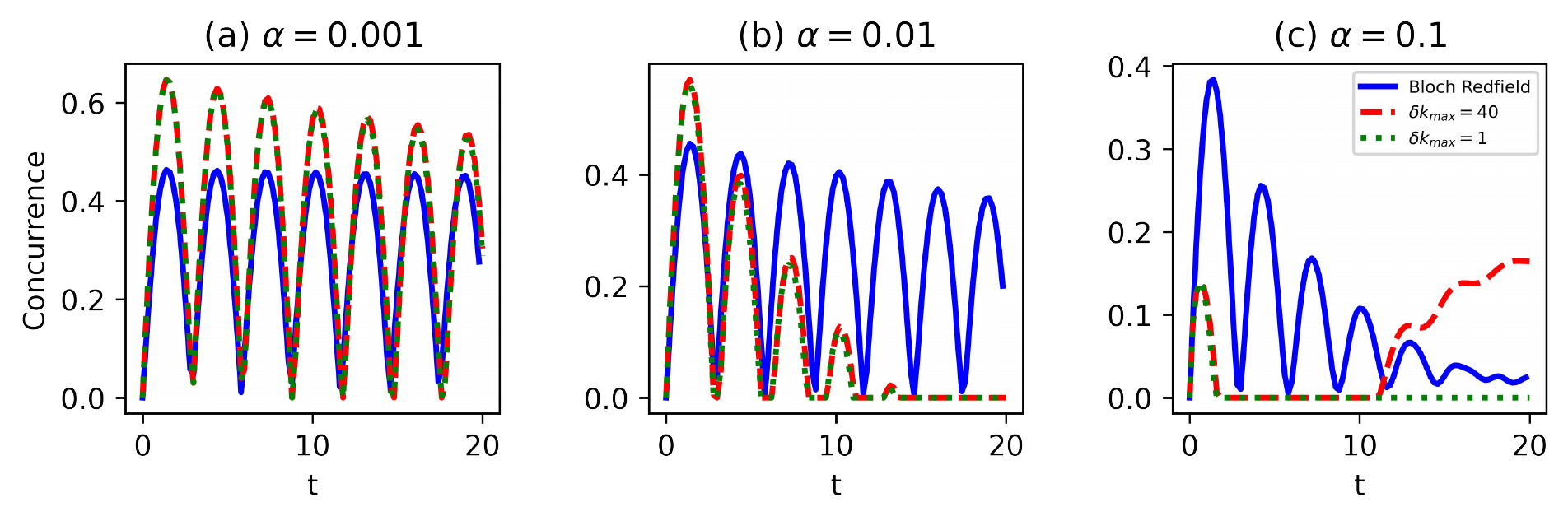}
\caption{\label{fig:BR} Entanglement vs. time in the Ohmic baths with different coupling strengths to the baths: (a) $\alpha=0.001$, (b) $\alpha=0.01$, and (c) $\alpha=0.1$. The red dashed line represents PT-TEBD with a memory cutoff $\delta k_{max}=40$, which captures the full non-Markovian effect. The green dotted line represents PT-TEBD with a memory cutoff $\delta k_{max}=1$, which is the Markov approximation used in this paper. The blue solid line represents the Bloch--Redfield theory under the Born--Markov approximation. The temperatures of the two baths are $T_1=T_2=0.2$. The system evolution starts from $|00\rangle$. The other parameters are $\omega_c=4$ and $\delta t=0.2$.}
\end{figure}

We now comment on the Markov approximation used in our paper. This approximation is similar to the classical Markov approximation, which means that the probability for a stochastic process to take the value $x_{n+1}$ at time $t_{n+1}$, under the approximation that it assumed values $x_i$ at previous times $t_i$, only depends on the last previous value $x_{n}$ at time $t_{n}$. When the memory cutoff $\delta k_{max}> 1$, the $\delta k_{max}$--rank ADT that encodes the evolution history of the system is propagated in QUAPI, and the quantum dissipative process can be viewed as the Markovian dynamics for an ADT \cite{IF2,IF3}. When $\delta k_{max}= 1$, the ADT degenerates into a density matrix. The Markov approximation used here is superior to the usual Bloch--Redfield approach. The main distinction is that the Born approximation, which is valid only up to order $g^2$ in the perturbation parameter, is not used here. Hence, the system-environment entanglement is preserved and contributes to the system's evolution. The second Markov approximation in the literature (assuming that the integral limits of time $t$ can be extended to $\infty$) is also not used here \cite{The_Theory_of_Open_Quantum_Systems,de2017dynamics}. 
In Fig. \ref{fig:BR}, we compare the results from different approaches, including the PT-TEBD algorithm with different cutoffs and the Bloch--Redfield master equation theory. The behaviors of the concurrences are very similar in the various approaches when the coupling strength is weak, $\alpha=0.001$, in Fig. \ref{fig:BR}(a). All of them oscillate and decay. The Bloch--Redfield master equation theory performs relatively well in this coupling regime. The trends overlap in PT-TEBD with different cutoffs, which alludes to the memory effect being insignificant in this coupling regime. The difference in results between these diverse approaches may come from the non-vanishing system-environment correlations. For an intermediate coupling strength, $\alpha=0.01$, the entanglement vanishes more quickly in PT-TEBD than in Bloch--Redfield theory. The Markov approximation used in PT-TEBD still works well in the intermediate coupling regime, $\alpha=0.01$, in Fig. \ref{fig:BR}(b). For a strong coupling strength, $\alpha=0.1$, the dynamics in PT-TEBD and Bloch--Redfield theory are dissimilar in Fig. \ref{fig:BR}(c). The Markovian approximation loses efficacy here. The three approaches give different predictions with a strong coupling strength. The entanglement is revived when the memory is fully conserved but dies permanently in the other two approaches. As we envisioned, the memory effect is significant in the strong coupling regime.

We discover that memory can boost the correlations at fixed times. However, more memory is not always better, and too much memory can also cause decoherence. This is illustrated in Appendix.\ref{memory and nonequilibrium}. We also examine how to confront decoherence from the environment in practical teleportation in Appendix.\ref{Engineering environment and teleportation}. The decay rates of entanglement and the fidelity of quantum teleportation can be slowed down drastically by regulating the external field.

\section{The dissipative dynamics of the Aubry-Andr\'{e} model\label{open MBL}}

\subsection{The evolution of the imbalance and the entanglement\label{open MBL dynamics}}

It has been shown that the entanglement in the localized phases of matter may persist for a long while \cite{wang2018time}. Hence, it is possible to protect the entanglement of the system by introducing the disorder into the environment. There are two questions to answer. First, most studies on the MBL/AL are concentrated on the closed system. It will be thermalized eventually if the system is open. However, can we still identify the ergodic phase and the MBL/AL phase, especially when the system strongly interacts with the environment? Second, can we protect the entanglement of the system by introducing the disorder into the environment? To answer the above question, we extend our model from a two-qubit system to a spin chain. Consider a finite spin-1/2 chain with open boundary conditions containing N sites. The Hamiltonian of the system is given by
\begin{equation}
H_{XXZ} =\sum_{i} J\left\{ S_{i}^{x}S_{i+1}^{x}+S_{i}^{y}S_{i+1}^{y}\right\}+ \Delta S_{i}^{z}S_{i+1}^{z}-\sum_{i}h_{i}S_{i}^{z}.
\end{equation}
When $\Delta \neq J$ and  $\Delta \neq 0$, it is a XXZ chain; when $\Delta=0$, the model is called the XX-spin chain model; when $\Delta=J$, it is called the Heisenberg chain. All of these models have a U(1) symmetry, the $\sum_{i}S_{i}^{z}$ is conserved. After the Jordan-Wigner transformation, the above spin-1/2 XXZ chain model is equivalent to the interacting spinless fermionic model,
\begin{equation}
H =\sum_{i} \frac{J}{2}\left\{ c_{i}^{\dagger}c_{i+1}+c_{i}c_{i+1}^{\dagger}\right\}+ \Delta \sum_{i}(n_{i}-\frac{1}{2})(n_{i+1}-\frac{1}{2})-\sum_{i}h_{i}n_{i},
\end{equation}
where $c_{i} (c_{i}^{\dagger})$ are creation (annihilation) operators, and $n=c_{i}^{\dagger}c_{i}$ are occupation number operators at site i. When the system is in a quasi-periodic potential $h_{i}=h\cos(2\pi\beta i)$ and $\Delta=0$, the model corresponds to the well-studied Aubry-Andr\'{e} model \cite{aubry1980analyticity}. Owing to the duality of the Aubry-Andr\'{e} model (as it retains the same form after a Fourier transformation), $h/J=1$ is the phase transition point. $h/J>1$ corresponds to the Anderson localization.  In the following, we set $J=1$ as the unit of energy. When $\Delta\neq 0$, it is an interacting Aubry-Andr\'{e} model, in which the MBL phase also emerges. There have been numerical results showing that the critical disorder is at $h_{c}\approx 2.4\pm0.25$ in such a model \cite{doggen2019many}.
The external environment influences the evolution of the chain system. We assume that there is a bath couple to the first site, the bath Hamiltonian plus the interaction Hamiltonian is
\begin{equation}\label{couple bath}
H_{bath}+H_{interaction} =\sum_{k} \omega_{k} a_{k}^{\dagger}a_{k} + S_1^{z}\sum_{k}  g_{k}a_{k}^{\dagger}+g_{k}^{*}a_{k},
\end{equation}

We mainly consider the environment with its Ohmic spectrum $J(\omega)=\sum_{k}|g_{k}|^2 \delta(\omega-\omega_K)=2\alpha\omega e^{-\frac{\omega}{\omega_{c}}}$ in this section. The spin number is $N=8$. We numerically simulate the dynamics of an initial state $\frac{1}{\sqrt{2}}(|11010101\rangle>+|01010100\rangle>)$, which means that the two ends of the spin chain are maximally entangled initially. A similar setting has been studied in the XXZ model \cite{wang2018time}. However, the AA model and the dissipative effect from the environment have been not considered. Let's first consider the spin chain isolated from the environment. To distinguish the phases of the chain, define the imbalance $\textit{I}(t)=\frac{N_{odd}(t)-N_{even}(t)}{N_{odd}(t)+N_{even}(t)}$. The experiments and theory have shown that the $\textit{I}(t)$ vanishes in the ergodic phase but keeps a non-vanishing value in the localization phases  \cite{schreiber2015observation,alet2018many}.
\begin{figure}[H]
\centering
\includegraphics[width=1\textwidth]{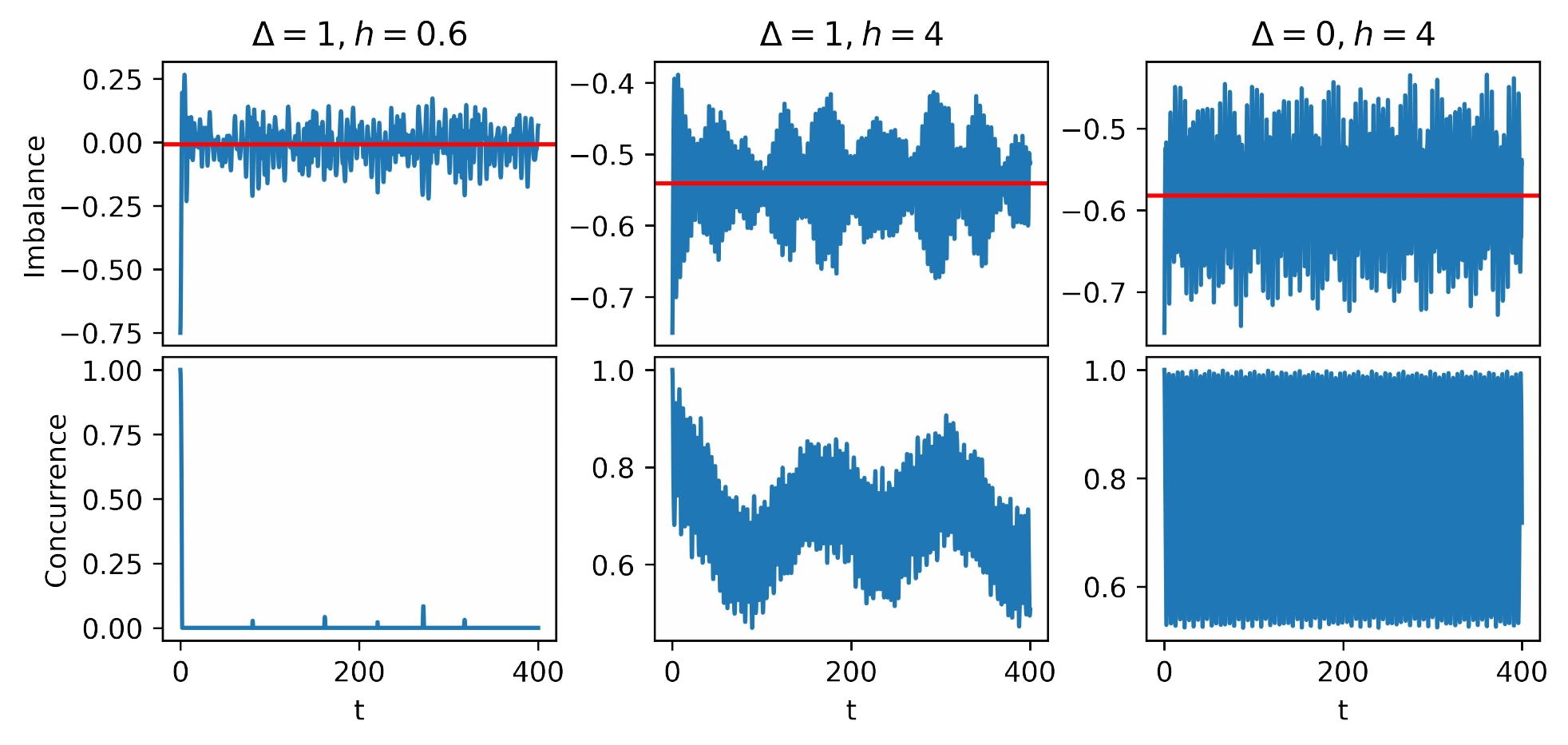}
\caption{\label{fig:without bath} Top panel: Imbalance dynamics in various phases. The red solid line is the average value.  Bottom Panel: Entanglement dynamics in various phases. Time step $\delta t=0.2$. Truncation error $\epsilon=10^{-6}$ for performing TEBD.}
\end{figure}
Restricted by computing power, our research can only concentrate on finite time zones. In Fig.\ref{fig:without bath}, we display the dynamics of imbalance for both the interacting and non-interacting cases. When $h=0.6$, the imbalance decays, rises, and falls around zero for $\Delta=1$, and the time average value is very close to zero. Therefore, the AA chain under these parameters is in the ergodic phase. When $h=4$, the time average of imbalance keeps a finite value and fluctuates around it. These confirm a localization phase, in which the system keeps a memory of the initial state. More concretely, it corresponds to the MBL phase when $\Delta=1$ and the AL phase when $\Delta=0$ respectively. The entanglement dynamics of the chain's ends are also computed. From the view of the open system theory, the ends serve as a two-qubit system, and the remaining qubits can be viewed as a high temperature environment. The concurrence declines to zero rapidly in the ergodic phase and also revives but soon dies again at some later times. The entanglement fluctuates around some finite value in the MBL/AL phase. Meanwhile, the entanglement exhibits a fluctuation behavior over a larger time scale in the MBL phase. The time average of the concurrence in AL is larger than that in MBL. The differences in imbalance and concurrence in ergodic and MBL/AL phases originate from the diffusion of a particle’s wave packet being absent in a disordered/quasi-disordered environment, implying the initial information is partially reserved.  In addition, the slow entanglement spreading in the MBL/AL phases is limited by a version of the Lieb–Robinson bound on the information light-cone and can be depicted via out-of-time-order correlations \cite{wang2018time,huang2017out}. The non-vanishing entanglement sheds light on applying the MBL/AL phase to store quantum information. 
\begin{figure}[H]
\centering
\includegraphics[width=1\textwidth]{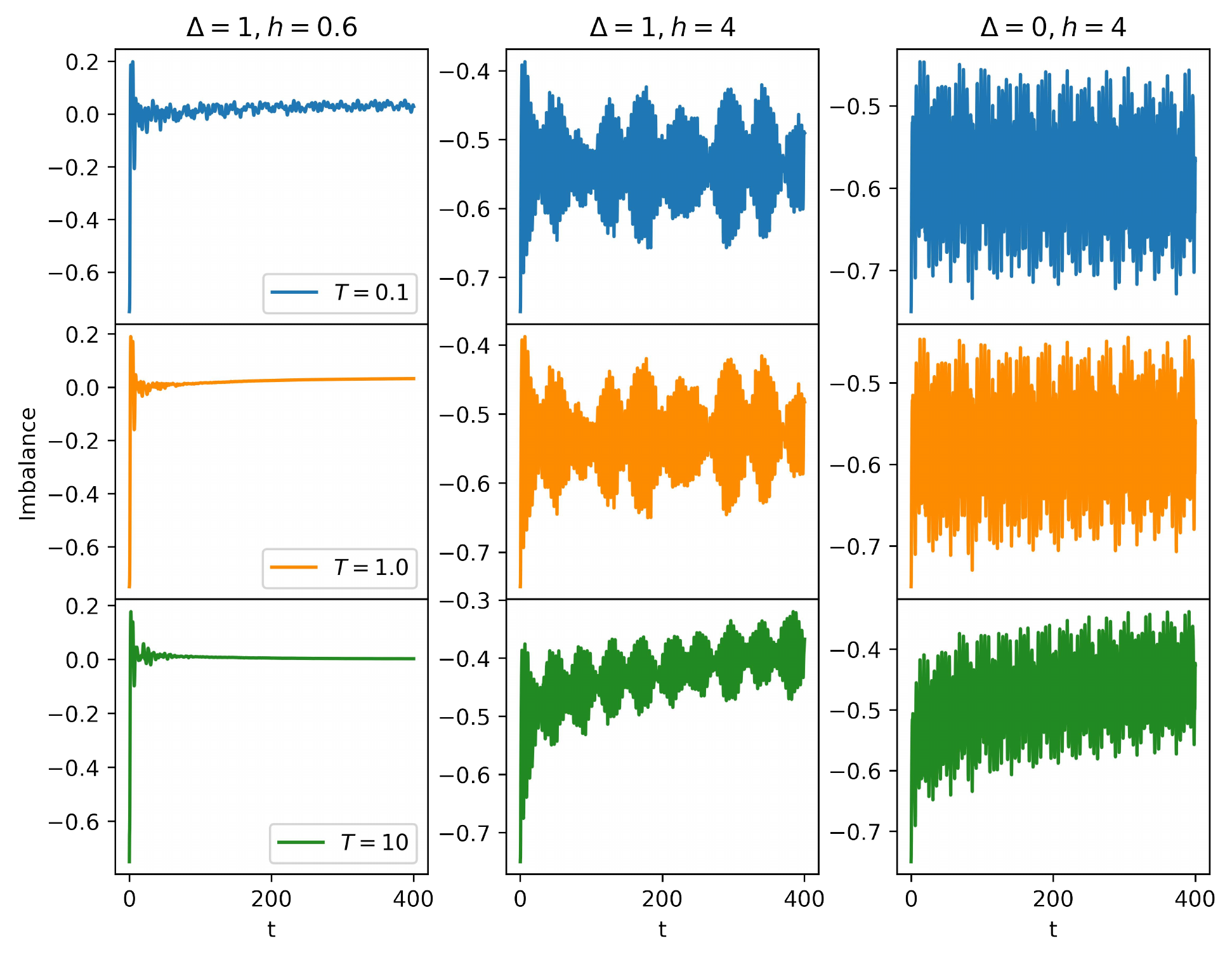}
\caption{\label{fig:with_bath imbalance} Imbalance v.s. time t when the first site in the AA chain couples with the Ohmic environment with different temperatures. The coupling strength is $\alpha=0.1$. Time step is $\delta t=0.2$. Memory cutoff is $\delta k_{max}=40$. Truncation error $\epsilon=10^{-5}$ for performing TEBD and $\xi=10^{-5}$ for deriving the process tensor.}
\end{figure}

We now study the dissipative dynamics of the AA chain, whose first site couples to an Ohmic bath. We found totally distinct imbalance dynamics in the ergodic phase or MBL/AL phase. The imbalance decays to zero rapidly in the ergodic phase. The higher the temperature is, the faster the decay becomes. This is shown in Fig.\ref{fig:with_bath imbalance} with parameters $\Delta=1$ and $h=0.6$. The imbalance dynamics is slightly influenced by the bath at low temperatures, for instance, $T=0.1 \& 1$, but changed significantly by the relatively high temperature $T=10$ in the MBL phase at parameters $\Delta=1$ and $h=4$. The thermalization effect is more significant at high temperatures, which aligns with intuition. These phenomena also take place in the AL phase at parameters $\Delta=0$ and $h=4$. We also observe that the imbalance dynamics in the MBL phase change more remarkably than that in the AL phase at $T=10$. One can anticipate that the AA chain in all cases will be thermalized at the long time limit, and the dissipation eventually eradicates localization. However, the systems with MBL/AL and ergodic Hamiltonians behave notably differently on the route to the steady state. This is consistent with \cite{levi2016robustness,fischer2016dynamics,medvedyeva2016influence,vakulchyk2018signatures} and hints that the imbalance in the early evolution can still be a reliable observable to identify localization or ergodic phase even when the system strongly interacts with the bath.

Why does the imbalance influenced by environment in the distinct phases behave differently and why is the AA chain in the MBL phase thermalized more easily than in the AL phase? We attempt to comprehend them through the lens of eigenenergy spectrum. Certain systems' eigenmodes resonate with the modes of the bath with the corresponding eigenenergy. As a consequence, the effect of the interaction is the most obvious at this moment, and so is the thermal flux from the bath. The resonance strength is also bounded by the spectrum density $J(\omega)$, preventing all of the system's eigenmodes from interacting with the bath. Taking the impact of the temperature into account, the thermal flux from the bath should be proportional to $J(\omega)n(\omega)$. In Fig.\ref{fig:thermal response} in Appendix.\ref{appendix c}, one can observe that of the three phases, the ergodic phase has the greatest number of eigenmodes that can resonate with bath modes, followed by the MBL phase and the AL phase. Meanwhile, the total flux contributed by the resonant response of all eigenenergy levels is the most significant in the ergodic phase, and that in the MBL phase ranks second and so on. As a result, the AA chain in the ergodic phase is most dramatically thermalized, while that in MBL phase ranks second.

\begin{figure}[H]
\centering
\includegraphics[width=1\textwidth]{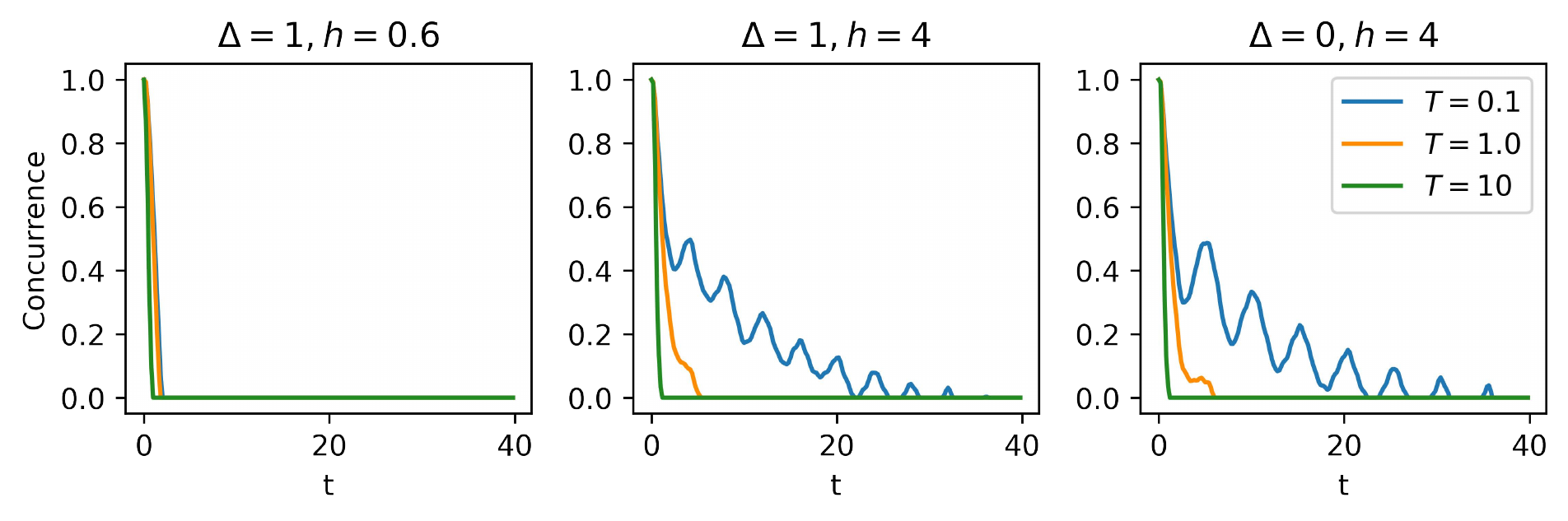}
\caption{\label{fig:with_bath entanglement} Entanglement dynamics when the first spin in the AA chain couples with the Ohmic environment with different temperatures. The parameters are similar with those in Fig.\ref{fig:with_bath imbalance}.}
\end{figure}

The entanglement dynamics between the AA chain's ends while one of the ends couples with a heat bath is computed in Fig.\ref{fig:with_bath entanglement}. The concurrence dramatically decreases to zero in the ergodic phase. Additionally, the descent accelerates upon increasing the temperature. There is no further revival in such a case. The concurrence damps oscillatorily when the bath is at low temperatures, for instance, $T=0.1$; yet, when the bath is at high temperatures in the MBL/AL phase, it also swiftly decreases to zero. This means that in environments at high temperatures, it is impossible to discern between different phases of entanglement evolution. Unlike imbalance dynamics, the detection of entanglement evolution can only be employed as an effective means to distinguish various phases in a transient time period even at low temperature. Why this happened?  First of all, the imbalance is the mean property of the whole chain and requires a longer relaxation time. Meanwhile, the entanglement will scramble to the environment drastically since the bath couples directly to one of the entangled ends. Thus, the entanglement can no longer be regarded as a global observable to detect the phase. To verify this point from the side, we also compute the entanglement evolution when the bath couples with the fourth spin in Fig.\ref{fig:with mid bath}. The concurrence takes a perceptible amount of time to decline to zero in the localization phase. Particularly, the existence of the bath only affects the entanglement evolution slightly in the AL phase. Since there is no direct interaction between the ends and the heat bath, the environment must first change the overall nature of the chain before it can further affect the entanglement of the chain's ends. To put it another way, the middle part of the quasi-disordered chain acts as a buffer layer to weaken the spreading of entanglement if the chain's ends serve as a two-qubit system. Does this mean that we can exploit the disordered environment as a buffer layer to protect the quantum correlation of the system?
\begin{figure}[H]
\centering
\includegraphics[width=1\textwidth]{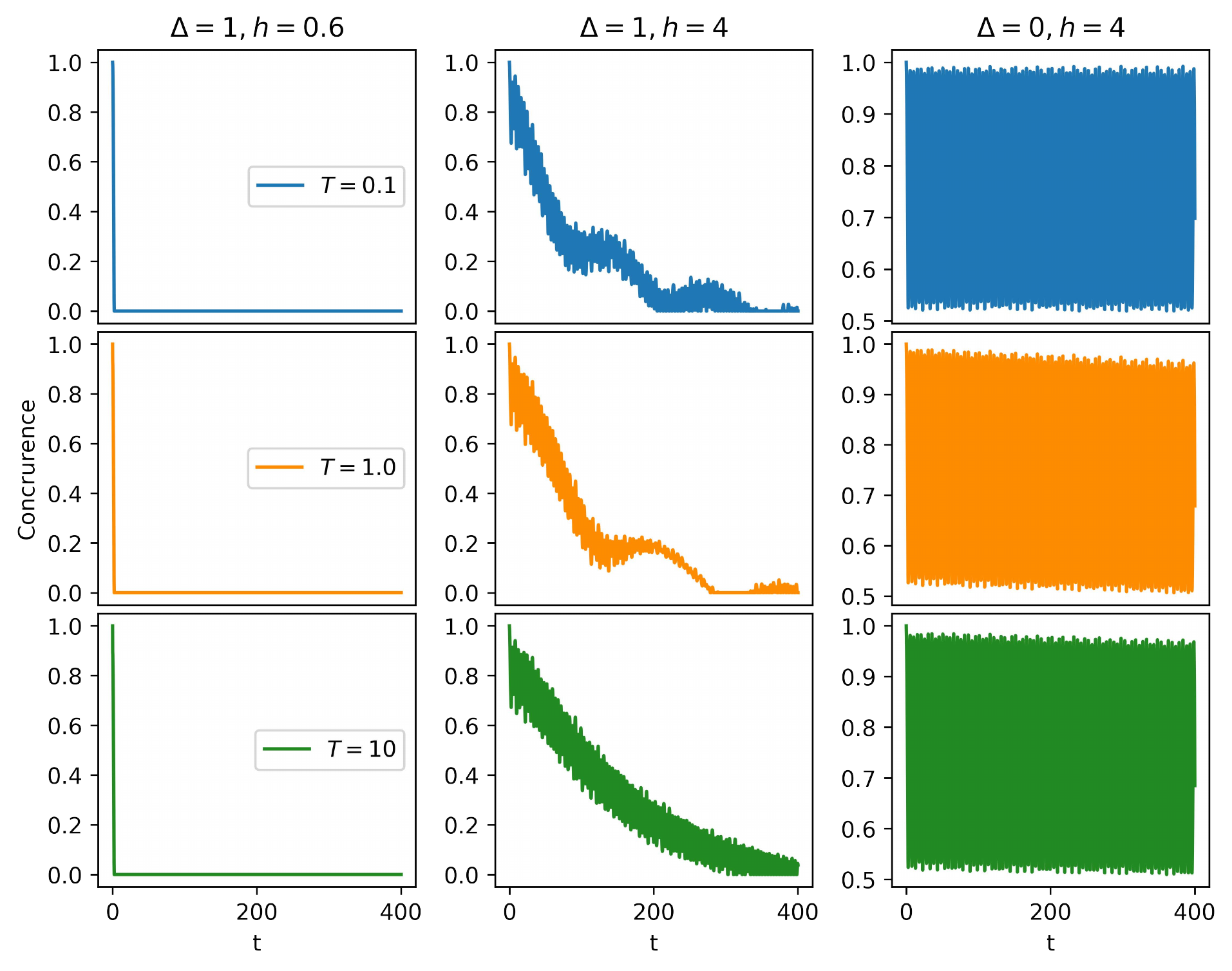}
\caption{\label{fig:with mid bath} Entanglement dynamics when the fourth spin in the AA chain couples with the Ohmic bath is at temperatures $T=0.1$, $T=1$, and $T=10$. The parameters are similar with those in Fig.\ref{fig:with_bath imbalance}.}
\end{figure}

\subsection{Entanglement protection: disordered environment as a buffer layer\label{buffer}}

To answer above question, we investigate the following scenario. There is a AA chain with eight sites, whose two ends are coupled to individual Ohmic baths. The fourth and the fifth site are maximally entangled, which can be regarded as the system that needs to be protected.  And other sites, which can be regarded as a protective layer, are all initially in the ground states. We compute the entanglement dynamics of the system. As a contrast, we simultaneously compute the entanglement evolution of the system without the protective layer, i.e., the system couples with the baths directly, as shown in Fig.\ref{fig:1}.
The system Hamiltonian in the unprotected case is $H_{S} =\sum_{i=4} J\left\{ S_{i}^{x}S_{i+1}^{x}+S_{i}^{y}S_{i+1}^{y}\right\}+ \Delta S_{i}^{z}S_{i+1}^{z}-\sum_{i=4,5}h_{i}S_{i}^{z}$, where $\Delta=1$ and $h_{i}=4\cos(2\pi\beta i)$. The bath Hamiltonian plus the interaction Hamiltonian for the single site is the same as Eq.\ref{couple bath}.

\begin{figure}[H]
\centering
\includegraphics[width=1\textwidth]{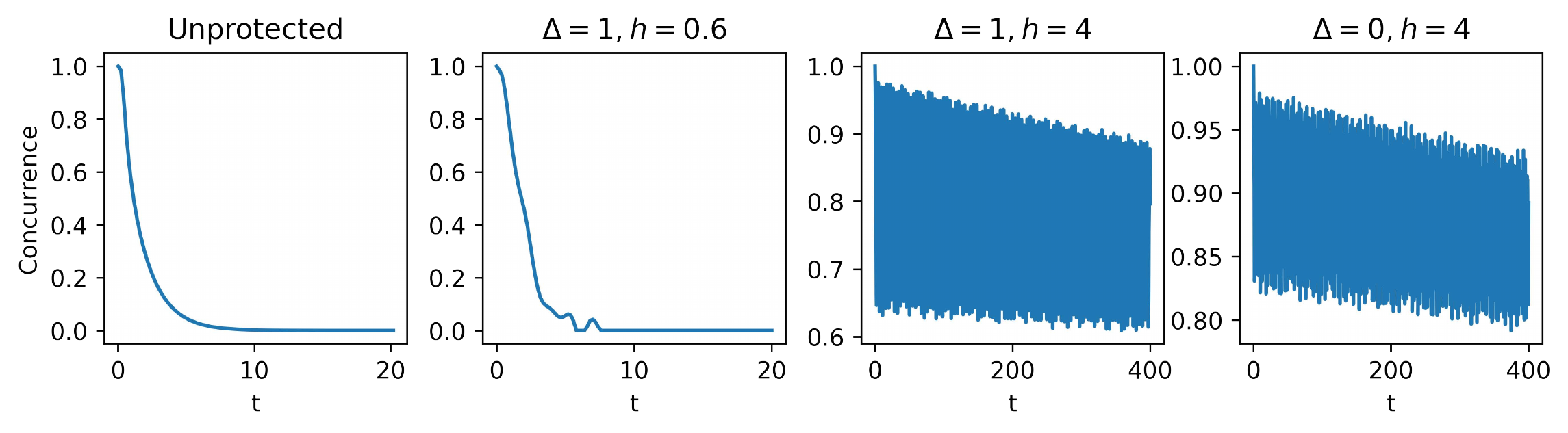}
\caption{\label{fig:buffer} Entanglement dynamics of the system. From left to right: without buffer; with ergodic buffer; with MBL buffer; with AL buffer. The temperatures of the baths are $T_1=0.8$ and $T_2=0.2$. Other parameters are similar with those in Fig.\ref{fig:with_bath imbalance}.}
\end{figure}

According to Fig.\ref{fig:buffer}, the entanglement in the bare system decays to zero in a remarkably brief period of time. Moreover, the buffer layer has no significant effect if it is in the ergodic phase. In contrast, the entanglement decays relatively slowly in a buffered system, where the buffer layer is modeled by the quasi-disordered spin chain in the MBL/AL phases. As we've seen, the AL buffer layer offers greater protection than the MBL buffer layer. This is also consistent with our earlier finding that the AA chains are more resistant to thermalization in the AL phase. We forecast that the entanglement of the system with the buffer disappearing after a considerable amount of time. This may offer a fresh concept for the storage of quantum information.

\section{Conclusion\label{Conclusion}}

In conclusion, we have used the numerically exact PT-TEMPO algorithm to study the quantum correlations within a two-qubit system in different types of reservoirs. Unlike for a system under the Markovian approximation, the memory influences the correlation dynamics significantly. Entanglement can re-occur after sudden death and persist for a long period of time. We find that environments with memory show a weak decoherence effect. The memory can improve the harvesting quantum correlations while reducing them depending on the lengths of the memories. For the various types of baths, the behaviors of the quantum correlations are somewhat distinct. The differences are mainly embodied in the duration and the amplitudes of the oscillations. Concretely, the quantum correlations of the two qubits oscillate for the longest time in the super-Ohmic baths while reaching a steady state the fastest in the sub-Ohmic baths. The super-Ohmic environment has the strongest memory effect. After a fixed significant period of time, the correlations can show non-monotonic behaviors with varying temperatures or temperature differences. Under certain conditions, nonequilibrium can boost the quantum correlations. This suggests a possible method of applying environmental engineering to promote or maintain the quantum correlations. We also compared the PT-TEBD under the Markovian approximation with the Bloch--Redfield master equation. The results for the dynamic behaviors show that the PT-TEBD under the Markovian approximation used in our paper exceeds the performance of the Bloch--Redfield master equation. The behaviors are similar when the system-bath coupling is weak. The differences may derive from the Born approximation used in the master equation, which omits system-environment correlations. The PT-TEBD under the Markovian approximation still performs well in the intermediate system-bath coupling regime, but the master equation does not. The dynamic behaviors are distinct when the system-bath coupling is strong and the Markovian approximation loses efficacy. We anticipate that only the PT-TEBD contains the total memory to provide an exact prediction.
%In a practical teleportation scenario, we have proposed a more efficient way to control the dissipation of the entanglement than adjusting the temperatures of the baths. Environmental engineering regulates the entanglement and fidelity of teleportation by introducing external field control. The decay rates of the entanglement and fidelity are lowered significantly. A more clever method to enhance entanglement and fidelity that combines environmental engineering and other technologies will be studied in future publications.

The AA chain at a specific initial state with bipartite entanglement, strongly coupled with a bath is also investigated. The dynamics of imbalance and entanglement of the ends are computed. When the AA chain is closed, the behaviors of the imbalance and the entanglement are totally different in the ergodic and the MBL/AL phase. The non-vanishing entanglement sheds light on applying the MBL/AL phase to store quantum information. Once there is a bath coupled with the first site of the chain, the imbalance disappears more rapidly in the ergodic phase. However, it is only at relatively high temperatures that the environment significantly changes the imbalance dynamics in localization phases. Predictably, in all cases, the AA chain will be thermalized over a long period of time, and dissipation will eventually destroy the localization. However, on the way to a steady state system with MBL/AL and an ergodic Hamiltonian, the behavior is markedly distinct. That means that the imbalance in the early evolution can still be a valid observable to differentiate localization or ergodic phases. We understand this in terms of whether the eigenmodes of the system can resonate with the environment and lead to energy exchange. The overall effect is that the response heat flux from the environment is the largest in the thermal phase, followed by the MBL phase. The entanglement of the ends will swiftly break up once there is a direct interaction between the ends of the chain and the bath. And when the environment is not directly coupled to the entangled ends but to intermediate sites, the entanglement of the chain's ends persists for a long time, which motivates us to take advantage of the disordered environment as a buffer to preserve quantum correlations. This was validated in the following comparisons.

\acknowledgments

He Wang thanks the TEMPO collaboration for introducing and explaining the use of the TEMPO algorithm.

\appendix

\section{The effect of nonequilibrium and memory on quantum correlations\label{memory and nonequilibrium}}
\begin{figure}[H]
\centering
\includegraphics[width=1\textwidth]{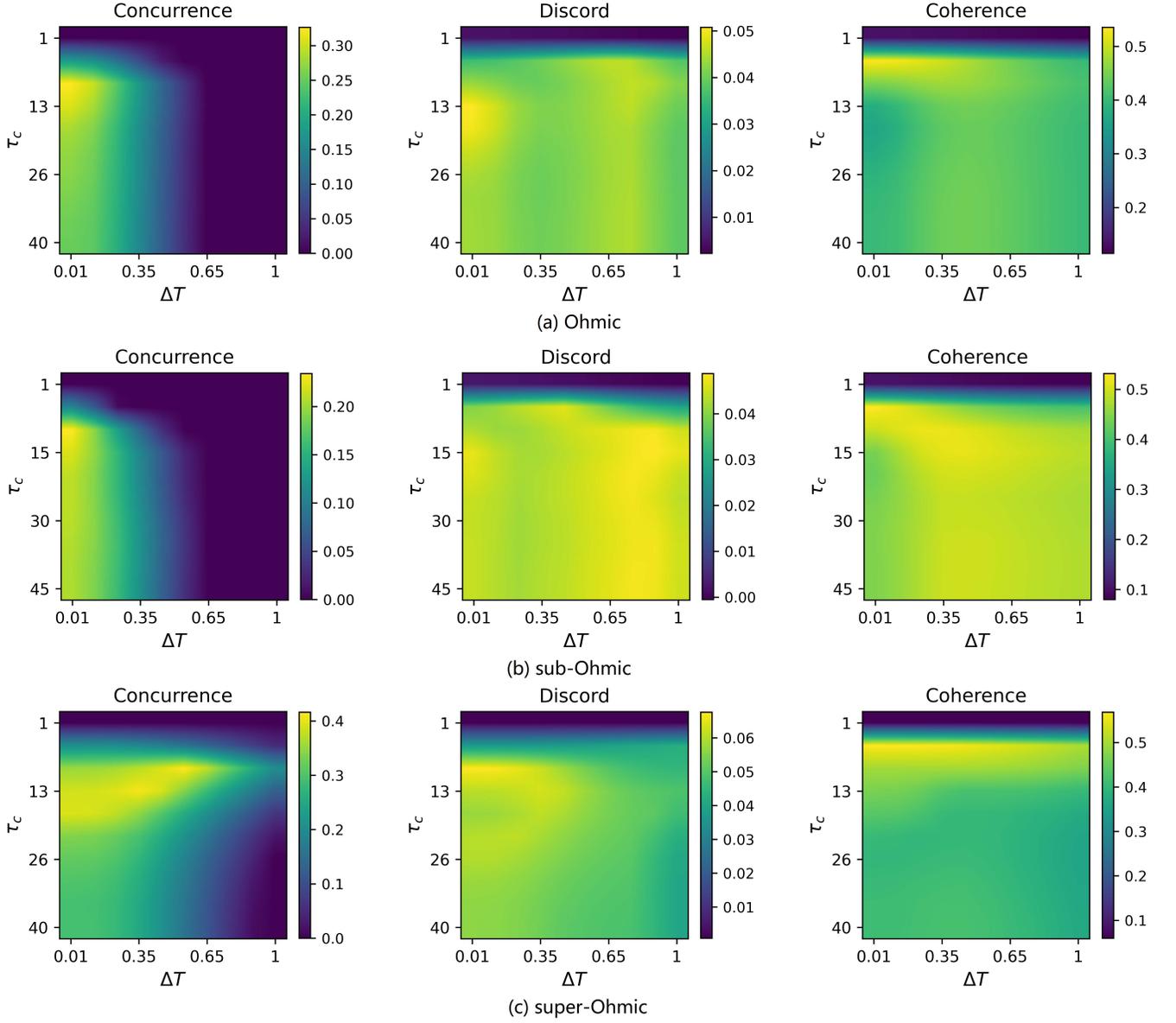}
\caption{\label{fig:Temperature_memory} Variations of the quantum correlations with respect to the memory cutoff and the temperature difference at the fixed time $t= 20$, (a) Ohmic, $\zeta=1$; (b) sub-Ohmic, $\zeta=0.6$; (c) super-Ohmic, $\zeta=2$. The temperature of one of the two baths is $0.01$, and the temperature of the other bath increases above $0.01$. The time step is $\delta t=0.2$ for both Ohmic and sub-Ohmic baths and $\delta t=0.1$ for super-Ohmic baths.}
\end{figure}

To see the influence of nonequilibrium and memory on the quantum correlations, we plot the variations of the quantum correlations with respect to the memory cutoff and the temperature difference at a fixed time $t=20$ in Fig. \ref{fig:Temperature_memory}. It is clear that the memory can improve the harvesting quantum correlations in all cases. In Fig. \ref{fig:Temperature_memory}, the correlations are seen to be enhanced by more memory, but they can also be lowered. This shows that more memory is not necessarily always better for improving quantum correlations. The entanglement varies monotonically with the temperature difference, and nonequilibrium may amplify the geometric discord and coherence in this case, as shown in Fig. \ref{fig:Temperature_memory}.

\section{Environmental engineering and teleportation \label{Engineering environment and teleportation}}

We observed the effects of nonequilibrium and memory on the system's evolution in the previous section, and this implies that we can employ environmental engineering to preserve or enhance the correlations. In some experiments, the temperatures and the system-environment couplings can both be relatively flexibly controlled, e.g., for transmon qubits coupled to a 1D transmission-line resonator in a circuit-QED system \cite{blais2021circuit}. Specific phenomena can be realized by fine-tuning the electromagnetic fields and other physical parameters. Actually, coherent control via periodic driving dubbed as Floquet engineering has become a versatile tool in quantum control \cite{oka2019floquet,bai2021quantum}. Inspired by this, we consider utilizing environmental engineering and external field control to influence the quantum correlations within the two qubits system. For the practical scenario of quantum teleportation, we will employ this kind of environmental engineering to enhance fidelity.   

\begin{figure}[H]
\centering
\includegraphics[width=0.8\textwidth]{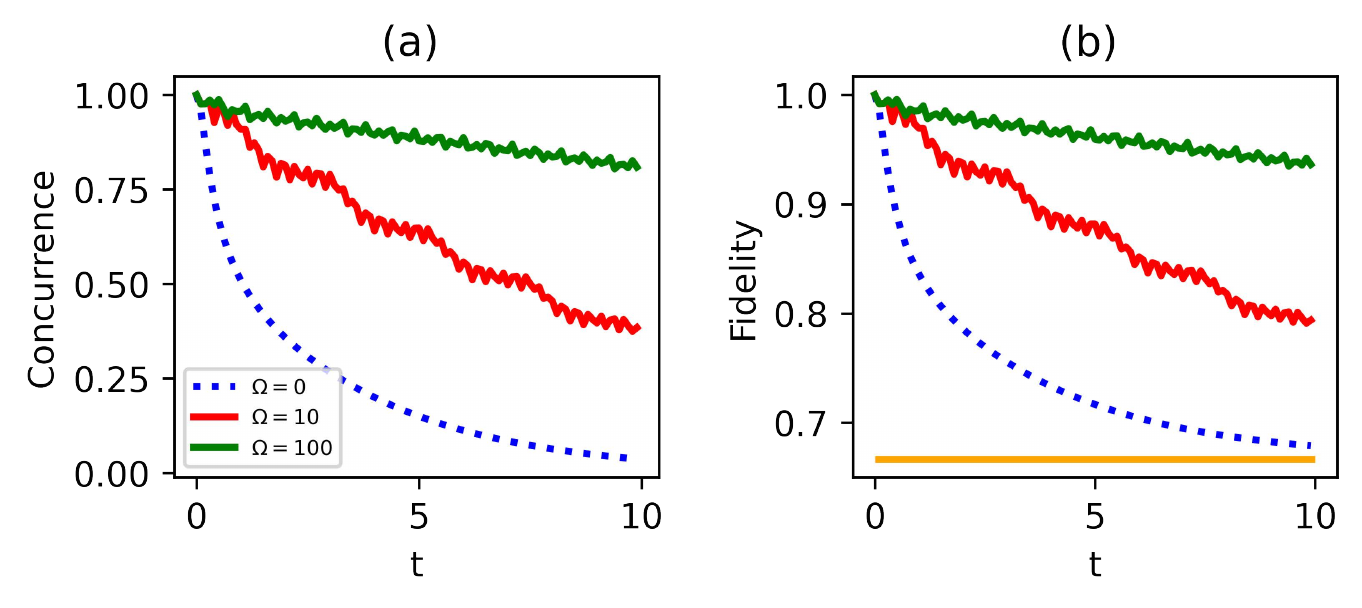}
\caption{\label{fig:Enigeering} (a) Entanglement and (b) fidelity vs. time in the Ohmic baths with different external field control strengths $\Delta(t)= \Lambda\sin(\Omega t)$. The blue dotted line represents $\Omega = 0$. The red solid line represents $\Omega = 10$. The green solid line represents $\Omega = 100$. The orange line is $\frac{2}{3}$, which is the upper bound for classical teleportation. The temperatures of the two baths are $T_1=T_2=0.2$. The other parameters are $\omega_c=5$, $\delta t=0.1$, $\alpha=0.1$, $\delta k_{max}=30$, and $\Lambda=50$. We use a Gaussian-type cutoff in the density spectrum here, i.e., $J(\omega)=2\alpha\omega \exp{(-\frac{\omega^2}{\omega_{c}^2})}$.}
\end{figure}
The separate initial maximally-entangled qubit pair suffer from the decoherence effect in the qubit distribution procedure. Certain powerful approaches have been proposed to weaken the decoherence effect and improve the fidelity, for example, weak quantum measurement (WM \cite{pramanik2013improving}) and environment-assisted measurement (EAM \cite{harraz2021protected}) technology. However, most of the quantum channels considered are simple phase damping channels, amplitude damping channels, etc., or their combinations. Only a few studies have considered the non-Markovian channel \cite{ban2005decoherence,motavallibashi2021non}. Here we propose a more complicated quantum channel. The total Hamiltonian of the single-qubit plus its bath is modeled as
\begin{equation}
\label{eq:18}
H_{total}=H_{S}+H_{SB}=H_{S}+H_{B}+H_{I}.
\end{equation}
The system Hamiltonian is $H_{S}=\frac{\omega}{2}\sigma_{z}+\frac{\Delta(t)}{2}\sigma_{x}$, where $\frac{\Delta(t)}{2}\sigma_{x}$ is induced by the controlled external field and $\Delta(t)= \Lambda\sin(\Omega t)$ in this study. A similar Hamiltonian has been studied theoretically and experimentally in quantum dot systems \cite{optimal_control,koppens2007universal}. The controlled external field is the heart of environmental engineering. The remaining part of the total Hamiltonian $H_{B}+H_{I}=\sum_{k}\omega_{k}\hat{a}_{k}^{\dagger}\hat{a}_{k}+\sigma_{z}\sum_{k}(g_{k}\hat{a}_{k}^{\dagger}+g_{k}^{*}\hat{a}_{k})$. There is no inter-qubit interaction in this case. The optimal fidelity of a general mixed state $\rho$ over all strategies is given as \cite{horodecki1996teleportation}

\begin{equation}
\label{eq:17}
F=\frac{1}{2}(1+\frac{1}{3}Tr(\sqrt{C^{\dagger}C})).
\end{equation}
where $C$ was already defined in Eq. \ref{eq:15}. Note that the state forming the quantum channel is useful for teleportation only when $F>\frac{2}{3}$, which is the upper bound for classical teleportation \cite{popescu1994bell}.

The system evolution starts from the maximally-entangled state $\frac{1}{\sqrt{2}}(|00\rangle+|11\rangle)$. In Fig. \ref{fig:Enigeering}, we plot the variation of the concurrence and fidelity with time and see that their behaviors are similar. When $\Omega=0$, there is no external field control, and both decrease monotonically. However, the concurrence and fidelity oscillate rapidly, and the decay trends are significantly reduced once we introduce the external field. Remarkably, the descending slope of the fidelity is minimal when $\Omega=100$. The introduced external field that regularly drives the system's dynamics competes with the thermal baths from which the dissipative dynamics originate. This can lead to the attenuation of the decoherence effect. Therefore, one can use a suitable modulated external field to prevent the entanglement from decaying quickly and hence enhance the fidelity. A more clever method to enhance entanglement and fidelity that combines environmental engineering and other technologies will be studied in future publications. % A more clever method is that combine the environmental engineering and WM or EAM technology. This will be studied in further publications.   

\section{The resonant response of the AA chain to the Ohmic environment\label{appendix c}}
\begin{figure}[H]
\centering
\includegraphics[width=1\textwidth]{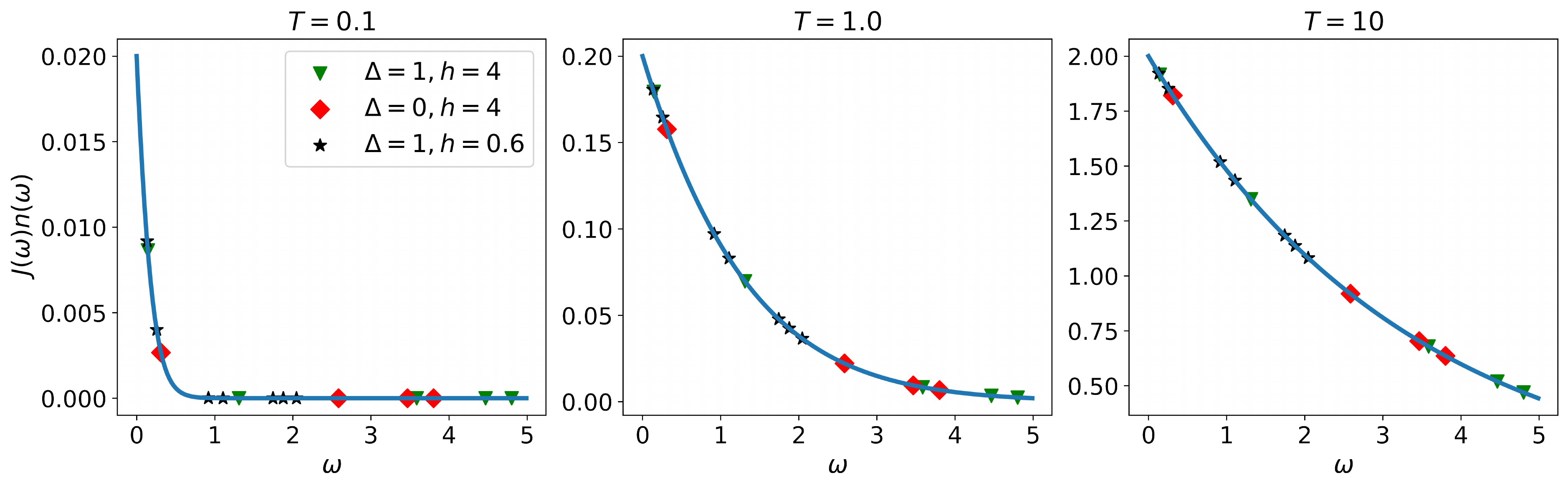}
\caption{\label{fig:thermal response} The variation of $J(\omega)n(\omega)$ with respect to $\omega$ at different temperatures. The green triangle, red diamond, and black star represent the eigenmodes in MBL, AL, and ergodic phase respectively, which can resonate with the corresponding modes in the bath. }
\end{figure}

In the main text, we argue that the thermal flux from the heat bath should be proportional to $J(\omega)n(\omega)$. $J(\omega)n(\omega)$ varies with $\omega$ at different temperatures are plotted in Fig.\ref{fig:thermal response}.  The various marks on the line represent the eigenmodes in MBL, AL, and ergodic phase respectively, which can resonate with the corresponding modes in the bath. One can observe that, of the three phases, the ergodic phase has the greatest number of eigenmodes that can resonate with bath modes, followed by the MBL phase and the AL phase. Meanwhile, the total flux contributed by the resonant response of all eigenenergy levels is the most significant in the ergodic phase, and in the MBL phase, it ranks second and so on. As a result, the AA chain is thermalized most dramatically in the ergodic phase, followed by the MBL phase.
\iffalse 
\section{The entanglement evolution when the bath couples with the fourth spin\label{appendix d}}
\begin{figure}[H]
\centering
\includegraphics[width=1\textwidth]{output/with mid bath_entanglement.pdf}
\caption{\label{fig:with mid bath} Entanglement dynamics when the fourth spin in the AA chain couples with the Ohmic bath is at temperature $T=0.1$. The parameters are similar with those in Fig.\ref{fig:with_bath imbalance}.}
\end{figure}
In Fig.\ref{fig:with mid bath}, we compute the entanglement evolution when the fourth spin in the AA chain couples with the Ohmic environment at temperature $T=0.1$. The concurrence still decays to zero drastically in the ergodic phase. The concurrence takes a perceptible amount of time to decline to zero in the MBL phase. While the existence of the bath only affects the entanglement evolution slightly in the AL phase. In this case, the evolution of entanglement can still be regarded as observable to identify distinct phases. We found that the entanglement can survive for a long time period in localization phases once there is no direct interaction between the ends and the heat bath. The environment must first change the overall nature of the chain before it can further affect the entanglement of the chain's ends. In other words, the middle part of the quasi-disordered chain acts as a buffer layer to weaken the spreading of entanglement. This inspires us to take advantage of disordered environments as buffers to protect quantum correlations.
\fi
\clearpage%%another page
\nocite{*}


\end{document}